\documentclass[
    aps,
    prb,
    preprint,
    superscriptaddress,
    amsmath,
    amssymb,
    longbibliography,
    floatfix
]{revtex4-2}

\usepackage[T1]{fontenc}
\usepackage[utf8]{inputenc}

\usepackage{graphicx}
\usepackage{float}

\usepackage{bm}
\usepackage{dcolumn}

\usepackage{xcolor}
\usepackage[
    colorlinks=true,
    linkcolor=blue,
    citecolor=blue,
    urlcolor=blue
]{hyperref}

\newcommand{\aSiOtwo}{a-SiO$_2$}

\newcommand{\SiOfour}{SiO$_4$}

\newcommand{\OSiO}{O--Si--O}

\newcommand{\SiO}{Si--O}

\newcommand{\OSiOmax}{\ensuremath{\mathrm{O}\!-\!\mathrm{Si}\!-\!\mathrm{O}_{\max}}}

\newcommand{\figref}[1]{Fig.~\ref{#1}}
\newcommand{\tabref}[1]{Table~\ref{#1}}
\newcommand{\secref}[1]{Sec.~\ref{#1}}

\begin{document}

\title{Statistically Robust Characterization of Polaron and Bipolaron Intrinsic Traps in Amorphous Silica Reveals the Role of Network Flexibility in Electron Localisation}

\author{Robin Delhomme}
\affiliation{Advanced Research Center for Nanolithography, Science Park 106, 1098 XG Amsterdam, The Netherlands}
\affiliation{Institute of Theoretical Physics, Institute of Physics, University of Amsterdam, Science Park 904, 1098 XH Amsterdam, The Netherlands}

\author{Jonathon Cottom}
\affiliation{Advanced Research Center for Nanolithography, Science Park 106, 1098 XG Amsterdam, The Netherlands}
\affiliation{Institute of Theoretical Physics, Institute of Physics, University of Amsterdam, Science Park 904, 1098 XH Amsterdam, The Netherlands}

\author{Emilia Olsson}
\affiliation{Advanced Research Center for Nanolithography, Science Park 106, 1098 XG Amsterdam, The Netherlands}
\affiliation{Institute of Theoretical Physics, Institute of Physics, University of Amsterdam, Science Park 904, 1098 XH Amsterdam, The Netherlands}

\date{\today}

\begin{abstract}
Intrinsic electron localization in amorphous silica (\aSiOtwo{}) contributes to charge trapping and defect generation in silica-based electronic materials – a phenomenon  that degrades logic-device performance but can be exploited for memory applications. However, the structural distribution of one- and two-electron traps remains incompletely characterized, largely because the disordered covalent network of \aSiOtwo{} spans a vast configurational space. To address this, we perform statistical sampling to select a representative 200-model subset of defect-free bulk \aSiOtwo{} structures from a structurally converged parent set of 2512 models, generated via classical melt--quench molecular dynamics simulations. We then characterize the intrinsic one- and two-electron traps in this subset using HSE06 density functional theory calculations. Analysis of neutral precursor sites shows that trapping propensity is not determined by a universal O--Si--O angle or Si--O bond length; it is controlled by the accessibility of the relaxed trap geometry, requiring both local strain and medium-range network flexibility. Single electrons predominantly form edge-centred polarons on distorted SiO$_4$ tetrahedra, with occupied states 3.1--4.3~eV below the conduction band minimum. Injection of a second electron drives stronger lattice relaxation, most commonly Si--O bond elongation leading to rupture, thereby forming threefold-coordinated-Si (Si$_3$) bipolarons and, in some cases, Si--Si-bonded bipolaron configurations. The calculated bipolaron energy ranges are consistent with optical absorption features at 3.7, 4.7, and 6.4~eV associated with intrinsic electron localization. 
\end{abstract}

\maketitle

\section{Introduction}
The Si/SiO$_2$ materials system has underpinned integrated electronics since the earliest metal-oxide-semiconductor (MOS) structures, and remains central to modern device reliability through the Si/a-SiO$_2$ interface, ultra-thin interfacial oxides, and SiO$_x$ networks \cite{Green2001UltrathinLimits,IEEEInternationalRoadmapforDevices2024InternationalEdition}. Continued scaling and architectural transitions, including FinFET and gate-all-around devices, sustain performance, but also make device behaviour increasingly sensitive to individual and correlated defects in gate dielectrics and interfacial layers \cite{IEEEInternationalRoadmapforDevices2024InternationalEdition}. Key reliability challenges in \aSiOtwo{} thin films include charge trapping, bias- and temperature-driven defect generation, stress-induced leakage current (SILC), and time-dependent dielectric breakdown (TDDB) \cite{Stathis2001PhysicalCircuits,Suehle2002UltrathinCharacterization,McPherson2012TimeRevisited}. Closely related field-assisted defect dynamics are also exploited functionally in SiO$_x$-based resistive switching memories, where electrically formed and ruptured conductive paths provide non-volatile resistance states \cite{LaTorraca2019MultiscaleMemory,Waser2007Nanoionics-basedMemories,Yao2010ResistiveOxide,Mehonic2018SiliconSwitching}. This duality, failure in logic and function in memory, motivates atomistic identification of the oxide traps and precursors that control both degradation and switching.

Electrically active defects in silicon MOS structures are commonly classified as: (i) interface states at Si/a-SiO$_2$ ($D_{it}$), (ii) near-interfacial border traps that exchange charge with the channel on experimental time scales, and (iii) deeper oxide traps that become accessible or are generated under high-field stress \cite{Lenahan1982EffectInterface,Fleetwood1992BorderDevices,FleetwoodPantelidesSchrimpf2008}. These populations are probed using complementary electrical and spectroscopic methods. Capacitance--voltage and conductance-based measurements, together with charge pumping, quantify interface-state densities and recombination at the interface \cite{Lenahan1982EffectInterface,Fleetwood1995BORDERRELIABILITY,J.S.Brugler1969ChargeDevices}. Deep-level transient spectroscopy (DLTS) and related transient methods provide information on trap energetics and capture/emission kinetics \cite{Lang1974Deep-levelSemiconductors}. Low-frequency noise and random telegraph noise (RTN) resolve single- and few-trap dynamics in scaled devices \cite{Fleetwood1992BorderDevices,Fleetwood1995BORDERRELIABILITY,Kirton1989NoiseNoise,Machlup1954NoiseSignal,Grasser2012StochasticInstabilities}, while electron paramagnetic resonance (EPR) and optical spectroscopy provide chemical fingerprints for selected defect families \cite{Lenahan1982EffectInterface,Lenahan1998WhatSystem}. Under elevated electric fields, trap creation is observed through SILC and ultimately TDDB, typically described using trap-assisted tunnelling and percolation frameworks \cite{DiMariaCartier1995,Kimura1996ConductionFilms,Sune2007ABreakdown,Stathis2001PhysicalCircuits,McPherson2012TimeRevisited}.

Despite this toolbox, two key gaps remain in our understanding of charge trapping in \aSiOtwo. Firstly, the known E$'$, oxygen-deficient centre (ODC), and non-bridging oxygen hole centre (NBOHC) families do not, by themselves, provide a complete atomistic inventory of the electrically active traps inferred from nanoscale-device RTN, time-dependent defect spectroscopy (TDDS), SILC, and breakdown measurements \cite{Skuja1998OpticallyDioxide,Afanasev2000ChargeSilicon,Skuja2011VisibleSiO2,Grasser2010Time-dependentTransistors,Grasser2012StochasticInstabilities,Campbell2009RandomNMOSFETs,Stampfer2020Semi-automatedTransistors}. Device-level measurements reveal broad distributions of trap kinetics and electrostatic coupling strengths, while high-field stress experiments show that native precursors can be supplemented by field-assisted defect generation and spatial correlation \cite{DiMariaCartier1995,Kimura1996ConductionFilms,Stathis2001PhysicalCircuits,Sune2007ABreakdown,LaTorraca2019MultiscaleMemory}. This motivates the search for structurally intrinsic trapping precursors in nominally defect-free a-SiO$_2$.

Secondly, the atomistic identity and statistical distribution of such intrinsic electron-trapping precursors remain incomplete. Computational and optical studies have proposed disorder-driven, polaronic electron localisation in amorphous silica, here referred to as intrinsic electron traps (IETs), associated with locally strained or flattened SiO$_4$ motifs \cite{El-Sayed2014Nature2,El-Sayed2015OpticalSiO2,Strand2018IntrinsicChallenges,Gao2016AInjection,Li2023TheoreticalSilica}. Low-temperature optical absorption measurements on high-purity, electron-irradiated silica reveal bands near 3.7, 4.7, and 6.4 eV that have been assigned, with support from atomistic calculations, to localised electrons at strained amorphous precursor sites \cite{El-Sayed2015OpticalSiO2}. However, these assignments remain debated because the bands overlap with features of established NBOHC and ODC families and lack a unique EPR fingerprint \cite{Skuja1998OpticallyDioxide,Skuja2011VisibleSiO2,Girard2019OverviewFibers,Griscom2011Trapped-electronSynthesis,Cannas2004VacuumSilica,Vaccaro2008LuminescenceLaser}. For comparison, some spectroscopic assignments are more securely established, including the $\sim$5.8 eV band of the E$'_\gamma$ centre and the $\sim$7.6 eV band associated with Si--Si bonds \cite{Agnello2008OpticalMeasurements,Hosono1991ExperimentalGlass}.

Early atomistic work by Bersuker \textit{et al.} linked electron trapping in SiO$_2$ to localised oxygen states on strained Si--O bonds \cite{Bersuker2002AStates}. Subsequent calculations by El-Sayed \textit{et al.} identified wide O--Si--O angles ($\geq 132^\circ$), together with elongated Si--O bonds, as structural fingerprints of IET precursor sites and estimated their concentration to be of order $4\times10^{19}$ cm$^{-3}$ \cite{El-Sayed2013IdentificationCalculations,El-Sayed2014Nature2}. In those calculations, electron trapping drives further widening of the O--Si--O angle towards a nearly linear configuration, elongates the neighbouring Si--O bonds, and produces a state several electronvolts below the conduction band edge. More recently, Li \textit{et al.} investigated strained amorphous silica cells and suggested that an IET precursor containing both a wide O--Si--O angle ($\geq 132^\circ$) and a long Si--O bond ($\geq 1.660$~\AA) can undergo electron-induced Si--O bond rupture \cite{Li2023TheoreticalSilica}.

However, previous studies have sampled only a limited subset of the structural environments available in amorphous silica. Because \aSiOtwo{} lacks a periodic lattice, nominally similar local motifs can be embedded in very different medium-range networks, which can alter their ability to relax, localise charge, and form stable traps. Existing IET precursor rules in the literature identify strained bond angles and bond lengths as trapping motif candidates \cite{Bersuker2002AStates,El-Sayed2013IdentificationCalculations,El-Sayed2014Nature2,El-Sayed2015OpticalSiO2,Li2023TheoreticalSilica}. Because such motifs lie in the upper tails of their respective distributions, attaining average structural agreement of models with experiment is not enough to establish transferable IET motif populations, concentrations and precursor criteria. A substantially larger sample is needed to capture a representative set of precursor candidates across the full range of medium-range environments in which they may occur. In addition, most existing work has emphasised one-electron IETs, whereas two-electron, bipolaronic IETs may be essential for understanding EPR-silent optical signatures, bond rupture, and the early stages of defect generation.

In this paper, we use multi-metric statistical convergence of amorphous network structure, 
to define a representative ensemble for intrinsic electron trapping in defect-free bulk \aSiOtwo{}. Starting from a set of 2512 amorphous structures obtained from classical melt--quench molecular dynamics simulations, we filter structures containing defects and select 200 models whose short- and medium-range geometric metrics reproduce the converged distributions of the remaining defect-free parent ensemble. These metrics collectively describe the geometry of the amorphous network, including local SiO$_4$ distortion, bond and angle strain, ring structure, and the separation and orientation of neighbouring SiO$_4$ tetrahedra. Electron-injection calculations on this geometrically converged ensemble then enable a systematic analysis of one- and two-electron IETs. This allows us to identify new IET species and determine how short- and medium-range order influence trap formation, trap depth below the conduction band edge, and the reliability of structural precursor criteria.

\section{Methods}
\label{sec:Methods}
\subsection{Computational workflow and model ensembles}
\label{subsec:workflow_model_ensembles}

The calculations were organised around three model ensembles. Classical melt--quench molecular dynamics (MD) generated candidate a-SiO$_2$ networks, which were subsequently refined with density functional theory (DFT) using the Perdew–Burke–Ernzerhof (PBE) functional \cite{Perdew1996GeneralizedSimple,*PhysRevLett.78.1396} and filtered to remove coordination defects. The resulting defect-free PBE ensemble was used to define the structural statistics of the glass and to select a representative subset for hybrid-DFT calculations. Electron trapping was then calculated with the Heyd--Scuseria--Ernzerhof HSE06 hybrid functional (HSE06) functional \cite{Heyd2003HybridPotential,*Heyd2006Erratum:8207,Heyd2004Assessmentfunctional} for neutral, one-electron-charged, and two-electron-charged cells.

A total of 2512 structures were optimised at the PBE level. Coordination filtering left 1865 defect-free models. Structural descriptors and representative-subset selection were based on this 1865-model ensemble, while HSE06 electron-trapping calculations were performed on a 200-model subset selected to reproduce the distribution of local and medium-range order.

The following python libraries were used for data analysis: Pandas \cite{Mckinney2010DataPython}, Numpy \cite{Harris2020ArrayNumPy}, Scipy \cite{Virtanen2020SciPyPython}, and Atomic Simulation Environment \cite{HjorthLarsen2017TheAtoms}. Matplotlib \cite{Hunter2007Matplotlib_A_2D_Graphics_Environment} was used for generating figures and Visual Molecular Dynamics 2.0 \cite{HUMP96} was used to visualise models. In all illustrations of atomic structures in this work, yellow spheres are used to represent silicon atoms and red spheres are used for oxygen atoms. 

\begin{table}[htbp]
    \centering
    \setlength{\tabcolsep}{10pt}
    \caption{Model ensembles used in this work.}
    \label{tab:model_ensembles_methods}
    \begin{tabular}{p{0.22\linewidth}p{0.10\linewidth}p{0.22\linewidth}p{0.28\linewidth}}
        \hline
        Ensemble & Number of models & Level of theory & Purpose \\
        \hline
        Candidate structures & 2512 & Classical MD followed by PBE & Initial structural ensemble \\
        Defect-free structures & 1865 & PBE & Structural statistics and representative subset selection \\
        Representative subset & 200 & HSE06 & Neutral, one-electron, and two-electron trapping calculations \\
        \hline
    \end{tabular}
\end{table}

\subsection{Generation and refinement of amorphous silica models}
\label{subsec:generation_refinement}

\subsubsection{Classical melt--quench molecular dynamics}
\label{subsubsec:melt_quench_md}

Amorphous silica models containing 192 atoms were generated with the LAMMPS molecular dynamics package \cite{thompson_lammps_2022}. Each cell contained 64 Si atoms and 128 O atoms, corresponding to 64 SiO$_2$ formula units, under three-dimensional periodic boundary conditions. All classical simulations used the NpT ensemble \cite{Parrinello1980CrystalStudy,Parrinello1981PolymorphicMethod} and a 0.5 fs timestep. Temperature and pressure were controlled using Nos\'e--Hoover dynamics \cite{Nose1984AMethods,Hoover1985CanonicalDistributions,Martyna1994ConstantAlgorithms} with a temperature damping of 100 time-steps (50 fs), and a pressure damping of 1000 time-steps (500 fs). The target pressure was held at 0~atm throughout the MD calculations. The Van Beest, Kramer and Van Santen (BKS) interaction potential \cite{VanBeest1990ForceCalculations} was used with the SHIK-1 parametrisation \cite{Sundararaman2018NewGlasses}, which combines the Buckingham and Coulomb pair potentials with a correction term to counteract unphysical short-range attractive interactions. The Buckingham part was implemented using the buck pair style with a cutoff of 8 \AA{}. To accelerate the calculations, the Coulomb and short-range correction terms were saved to a table that contained the energies and forces for each pair of atomic species for interactomic separations between 0.1 \AA{} to 10.00995 \AA{} in steps of 0.00005 \AA{}. The table pair style was used such that the energy and forces for a given interatomic separation were obtained by linear interpolation from the closest entries in the table.

Amorphous models were generated using a melt--quench procedure: (i) randomly place 64 Si and 128 O atoms in a cubic cell of side length 21 \AA{} and perform a conjugate-gradient energy minimisation with fixed cell geometry; (ii) heat from 1 K to 5000 K over 5 ps; (iii) hold the melt at 5000 K for 50 ps, allowing the volume and density of the cell to equilibrate; (iv) continue holding the melt at 5000 K while saving snapshots every 10 ps; (v) quench each snapshot from 5000 K to 300 K at 1 K ps$^{-1}$; (vi) equilibrate at 300 K for 50 ps; and (vii) minimise the final structure by conjugate-gradient energy minimisation with fixed cell geometry. The energy tolerance for the conjugate-gradient minimisations was set to $10^{-6}$ eV, and the force tolerance was $10^{-9}$ eV/\AA{}. The 10 ps interval between snapshots was chosen to reduce structural correlations between successive models; the corresponding decorrelation analysis is given in Sec. S1 A 2 of the Supplemental Material (SM). The total simulation time for the melt and equilibration of the melt was 25,175 ps; snapshots were generated until a sufficient number of models was acquired to converge the properties of interest. The convergence process is described in \secref{subsec:statistical_significance} and resulted in 2512 SHIK-1 models being generated. A detailed structural analysis of these models can be found in \cite{vitriflow_paper}. 

\subsubsection{PBE refinement and coordination filtering}
\label{subsubsec:pbe_refinement_filtering}

The classically quenched structures were refined with the Quickstep \cite{Vandevondele2005Quickstep:Approach} module of CP2K \cite{Hutter2014Cp2k:Systems,Kuhne2020CP2K:Calculations,Iannuzzi2026TheSimple}. The PBE functional was used with the DZVP-MOLOPT-SR-GTH basis set \cite{VandeVondele2007GaussianPhases} and Goedecker-Teter-Hutter pseudopotential \cite{Goedecker1996SeparablePseudopotentials,Hartwigsen1998RelativisticRn,Krack2005PseudopotentialsFunctionals}. The plane-wave cutoff was 550 Ry, selected to converge the energy to within $10^{-3}$~eV/atom. All calculations were performed at the $\Gamma$ point and the self-consistent-field convergence criterion was $10^{-6}$. Both the atoms' positions and the cell lengths were optimised using the BFGS minimiser, while the cell angles were maintained at 90$^\circ$. The convergence criteria are listed in Sec. S1 A 3 of the SM.

Coordination defects were removed before analysing electron trapping. In a 192-atom cell, a single coordination defect would correspond to an artificially high defect concentration and would introduce local strain not representative of defect-free bulk a-SiO$_2$. Si--O neighbours were identified with a first-shell cutoff of 2.40 \AA{}, chosen to ensure that all coordination defects were correctly identified and removed, as described in Sec. S1 A 4 of the SM.

The radial distribution function averaged over the 2512 PBE-optimised models had maxima at 1.63 \AA{} for Si--O, 2.66 \AA{} for O--O, and 3.16 \AA{} for Si--Si (Fig. S2 in the SM), in good agreement with experiment \cite{Warren1938THESTUDIES, Neuefeind1996BondSilica}. After coordination filtering, 1865 defect-free models remained, \textit{i.e.} with every Si atom four-fold coordinated and every O atom two-fold coordinated, with a maximum Si-O bond length of 1.71 \AA{} . Their densities ranged from 2.10 to 2.48 g cm$^{-3}$ (Fig. S3(e) in the SM), with an average of 2.26 g cm$^{-3}$, consistent with experimental values for silica glass \cite{BRUCKNER1970123}. Full distributions of Si--O bond lengths, O--Si--O angles, and Si--O--Si angles are reported in Figs. S3(a)-S3(c) in the SM.

\subsection{Structural descriptors of the neutral network}
\label{subsec:structural_descriptors}

\subsubsection{Bond network and candidate O--Si--O sites}
\label{subsubsec:bond_network_candidate_sites}

Structural descriptors of neutral, defect-free, cells were evaluated on the Si--O network, with the same 2.4~\AA{} Si--O cutoff used for coordination filtering. The Si--O RDF of such cells plateaus at 0 over the 1.72-2.42~\AA{} range. Therefore, any cutoff value in this range would yield the same structural description. O--O and Si--Si pairs were not treated as bonds in the pristine network. In contrast, charged cells contain defects which lead to non-zero RDF values in the 1.72-2.42~\AA{ range. A cutoff of 2.03~\AA{} was used for these cells, based on the \SiO{} bond rupture analysis in \secref{subsec:Ed_vs_Si3}. O--O pairs were not treated as bonds in charged cells, while Si--Si pairs were treated as bonds if they participated in an electron trap belonging to a distinctive trap depth distribution, as described in \secref{subsec:trap_identification_localisation_methods}.

A candidate precursor site was defined as an O--Si--O angle centred on a four-fold-coordinated Si atom. Each defect-free 192-atom model contains 64 SiO$_4$ units and therefore 384 O--Si--O candidate sites. These sites are used for the packing analysis, trap-site assignment, and precursor-score ranking.

\subsubsection{Tetrahedral distortion}
\label{subsubsec:tetrahedral_distortion_methods}

The distortion of a SiO$_4$ unit was measured relative to an ideal Si-centred tetrahedron with four Si--O bonds of length $l_{\mathrm{ref}}=1.62$ \AA{} and six O--Si--O angles of $\alpha_{\mathrm{ref}}=109.5^\circ$. The bond-length distortion ($\sigma_{\mathrm{BL}}$) was defined as
\begin{equation}
    \sigma_{\mathrm{BL}} =
    \sqrt{
    \frac{1}{4}
    \sum_{i=1}^{4}
    \left(l_i-l_{\mathrm{ref}}\right)^2
    },
    \label{eq:sigma_bl}
\end{equation}
where $l_i$ is the length of the $i$th Si--O bond. The bond-angle distortion ($\sigma_{\mathrm{BA}}$) was defined as
\begin{equation}
    \sigma_{\mathrm{BA}} =
    \sqrt{
    \frac{1}{6}
    \sum_{i=1}^{6}
    \left(\alpha_i-\alpha_{\mathrm{ref}}\right)^2
    },
    \label{eq:sigma_ba}
\end{equation}
where $\alpha_i$ is the value of the $i$th O--Si--O angle.

\subsubsection{Packing of neighbouring SiO$_4$ units}
\label{subsubsec:packing_descriptors_methods}

The local packing of neighbouring SiO$_4$ units was characterised for each O--Si--O candidate site, as illustrated in \figref{fig:updated_MRO_characterisation}(a). For a central angle O$_1$--Si$_0$--O$_2$, the two O atoms are bridging oxygens connected to neighbouring Si atoms, denoted Si$_1$ and Si$_2$. The separation of the neighbouring tetrahedra was measured by the Si second-nearest-neighbour distance,
\begin{equation}
    d_{\mathrm{Si2nn}} =
    \left|\mathbf{r}_{\mathrm{Si}_1}-\mathbf{r}_{\mathrm{Si}_2}\right|_{\mathrm{MIC}},
    \label{eq:d_2nn}
\end{equation}
where MIC denotes the minimum-image convention.

The relative orientation of the neighbouring tetrahedra was measured by the angle between the two neighbouring Si--O bond vectors that connect Si$_1$ and Si$_2$ to the oxygens of the central angle,
\begin{equation}
    \xi =
    \cos^{-1}\left(\hat{\mathbf{u}}_1 \cdot \hat{\mathbf{u}}_2\right),
    \label{eq:theta_nno}
\end{equation}
with
\begin{equation}
    \hat{\mathbf{u}}_1 =
    \frac{\mathbf{r}_{\mathrm{O}_1}-\mathbf{r}_{\mathrm{Si}_1}}
    {\left|\mathbf{r}_{\mathrm{O}_1}-\mathbf{r}_{\mathrm{Si}_1}\right|},
    \qquad
    \hat{\mathbf{u}}_2 =
    \frac{\mathbf{r}_{\mathrm{O}_2}-\mathbf{r}_{\mathrm{Si}_2}}
    {\left|\mathbf{r}_{\mathrm{O}_2}-\mathbf{r}_{\mathrm{Si}_2}\right|}.
    \label{eq:theta_nno_vectors}
\end{equation}
The joint distribution of $d_{\mathrm{Si2nn}}$ and $\xi$ is referred to as the SiO$_4$ packing distribution.

\begin{figure}[htbp]
    \centering
    \includegraphics[width=0.5\linewidth]{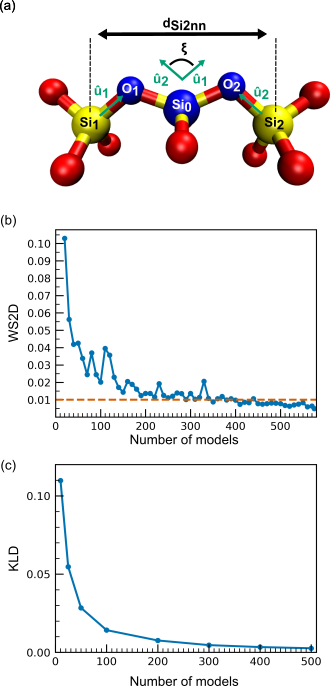}
    \caption{(a) Geometrical descriptors used to characterise the packing of neighbouring SiO$_4$ units around an \OSiO{} site. Yellow spheres are Si atoms, red spheres are O atoms, blue spheres are the atoms forming the \OSiO{} angle under consideration. The Si second-nearest-neighbour distance $d_{\mathrm{Si2nn}}$ measures the separation of the two tetrahedra adjacent to a central O--Si--O angle, while the angle $\xi$ measures their relative orientation. (b) Convergence of the high-strain region of the packing distribution in the defect-free structures. The orange dashed line shows the convergence threshold. (c) Kullback-Leibler divergence of the subset packing distribution relative to the global packing distribution of the defect-free structures, as a function of subset size.}
    \label{fig:updated_MRO_characterisation}
\end{figure}

\subsubsection{Ring topology}
\label{subsubsec:ring_topology_methods}

Ring topology was assigned on the Si--O bond graph. Atoms were graph nodes and Si--O bonds were graph edges. For each O--Si--O and Si--O--Si angle, the assigned ring was the shortest closed loop containing the two bonds that define the angle. If the angle contained more than one loop of the same minimum length, all such loops were counted. In this paper, ring lengths are reported in numbers of Si-O pairs, \textit{i.e.} a two-membered ring contains 4 atoms. 

Periodic boundary conditions were treated by tracking the accumulated cell translation around each candidate loop. Loops with a non-zero net translation in any cell direction were discarded, because they correspond to periodic strings rather than closed rings in the network. The graph-construction procedure was implemented with the NetworkX python library \cite{Hagberghagberg2008ExploringNetworkX}. The procedure details are given in Sec. S1 B 1 of the SM, along with the final ring length distribution in Fig. S3(f) and an example of an excluded periodic string in Fig. S4(b).

\subsubsection{Additional descriptors for precursor analysis}
\label{subsubsec:additional_precursor_descriptors_methods}

The precursor analysis used structural descriptors evaluated for each neutral O--Si--O candidate site. In addition to the O--Si--O angle, the precursor site analysis uses the following 5 descriptors: the average Si--O bond length associated with the candidate angle, the Voronoi volume of the central Si atom, the Si second-nearest-neighbour distance $d_{\mathrm{Si2nn}}$, and the maximum Si nearest-neighbour distance around the candidate site, defined as:

\begin{equation}
    d_{{\mathrm{Sinn}}}^{\mathrm{max}} = \max(\left|\mathbf{r}_{\mathrm{Si}_1}-\mathbf{r}_{\mathrm{Si}_0}\right|_{\mathrm{MIC}},\left|\mathbf{r}_{\mathrm{Si}_2}-\mathbf{r}_{\mathrm{Si}_0}\right|_{\mathrm{MIC}}).
    \label{eq:max_Sinn}
\end{equation}

Voronoi tessellation was performed using Scipy Voronoi and the volumes were calculated using Scipy ConvexHull. Periodic boundaries were handled by replicating the cell in the positive and negative x,y and z directions, and extracting the Voronoi volumes of the atoms in the central cell. 

\subsection{Ensuring statistical significance}
\label{subsec:statistical_significance}
Intrinsic electron trapping in amorphous silica is controlled by a combination of local strain and medium-range network flexibility. As such, while a study based on average agreement of RDFs, bond length and bond angle distributions is enough to identify possible trapping motifs, it cannot establish transferable precursor rules, trapping motif populations or concentrations. A rigorous representative sampling of the structural motifs relevant to electron trapping is a prerequisite for a robust characterisation of IET precursors.

The convergence of the packing distribution with respect to the number of models was first assessed using the Wasserstein 2-distance (WS2D), calculated using the Python Optimal Transport library \cite{flamary2021pot,flamary_2025_17176920}. The two dimensions of the packing distribution were min-max normalised to give them equal weight in the convergence process and to make the WS2D unitless. In 2D space, where both dimensions range from 0 to 1, the WS2D ranges between 0 and $\sqrt{2}$, we therefore consider 0.01 to be a good level of convergence. Figure \ref{fig:updated_MRO_characterisation}(b) shows the convergence of the high-strain region of the packing distribution, based on sites with \OSiO{} $> 123^{\circ}$. Adding models in batches of 10, the WS2D relative to the previous iteration's distribution falls below 0.01 after 500 models. Ultimately, 1865 defect-free models, of which 575 contained at least one \OSiO{} $> 123^{\circ}$, were generated to satisfy the convergence of the packing distribution.

HSE06 electron-trapping calculations were performed on a representative subset of the 1865 defect-free PBE-refined models. The subset was selected to reproduce the SiO$_4$ packing distribution of the full ensemble, since the aim is to relate electron trapping to local and medium-range structural order.

The ideal downsampled subset size was evaluated using the Kullback-Leibler divergence (KLD), defined as: 
\begin{equation}
    D_{\mathrm{KL}}(P\|Q) =
    \sum_i P(x)\log\left[\frac{P(x)}{Q(x)}\right],
    \label{eq:dkl}
\end{equation}
where $P$ is the kernel-density estimation (KDE) of the subset packing distribution, $Q$ is the KDE of the full defect-free packing distribution, and $x$ is a point on the two-dimensional space of the packing distribution. The KDEs where obtained using scikit-learn's KernelDensity class \cite{PedregosaFABIANPEDREGOSA2011Scikit-learn:Perrot} with a Gaussian kernel. The same bandwidth was applied to P and Q, which was calculated using the Silverman criterion on Q. The KLD was calculated using the Scipy.stats entropy function. Figure \ref{fig:updated_MRO_characterisation}(c) was obtained by repeating the sampling 100 times for each subset sample size and plotting only the smallest KLD. Following this method, a 200-model subset gave a suitable compromise between representativeness and HSE06 computational cost. 

The final HSE06 dataset was the 200-model subset with the smallest $D_{\mathrm{KL}}$ among 500 random draws. The inclusion of models with outlier sites of interest into the 200-model subset was manually enforced, as detailed in Sec. S1 C of the SM. The KLD of the final 200 model subset's packing distribution relative to the global packing distribution was 0.01.

\subsection{Electron-trapping calculations}
\label{subsec:hse_electron_trapping_methods}

The 200 selected defect-free models were treated with the HSE06 hybrid functional. The auxiliary density matrix method was used with the pFIT3 auxiliary basis set \cite{Guidon2010AuxiliaryCalculations} to reduce the cost of exact exchange calculations. The plane-wave cutoff was increased to 650 Ry, which converged the energy to within $10^{-6}$~eV/atom. The fraction of exact exchange was 25\%, and the short-range screening parameter was 0.11 bohr$^{-1}$. The self-consistent-field convergence criterion was $10^{-6}$, and all calculations were performed at the $\Gamma$ point.

Neutral cells were re-optimised at the HSE06 level before their electronic structures were calculated. Both the atoms' positions and the cell lengths were optimised using the BFGS minimiser, while the cell angles were maintained at 90$^\circ$. The convergence criteria are listed in Sec S1 D of the SM. The band gap of the neutral HSE06-optimised models ranged from 8.0 eV to 9.0 eV, with an average of 8.7 eV, in agreement with experiment \cite{Astasauskas2020OpticalSpectroscopy}.

One-electron traps were generated by adding one electron to each neutral HSE06-optimised cell, giving charge state $q=-1$. CP2K compensates the added charge with a uniform neutralising background. After electron injection, the geometry was relaxed to allow local structural response at the trap site.

Two-electron traps were generated by adding a second electron to the relaxed q = -1 geometry and relaxing the geometry again. The cell dimensions of q = -1 and q = -2 models were always kept constant to match those of the neutral models.

\subsection{Charged-cell corrections and energy definitions}
\label{subsec:charged_cell_corrections_methods}

Charged periodic calculations contain finite-size electrostatic errors because the excess charge interacts with its periodic images and the compensating background. These errors were corrected using the Lany--Zunger scheme \cite{Lany2009AccurateCalculations} with a static dielectric constant of 3.88 \cite{WhitmanDATAQUARTZ}. The resulting finite-size corrections to the total energies were on average $0.27$ eV for $q=-1$ cells and $1.09$ eV for $q=-2$ cells. 

CP2K sets the cell-averaged Hartree potential to zero. This convention can introduce an arbitrary potential offset when injecting charge in a cell. Potential alignment was therefore performed by following the method described in \cite{Durrant2018RelationConditions}. For our models, the average Hartree-potential shift was calculated in the 2.6 \AA{} region furthest from the trap site. The average potential-alignment correction was $+0.12$ eV for $q=-1$ cells and $+0.43$ eV for $q=-2$ cells.

Trapping energies were defined as the total energy difference of the charged cells before and after geometric relaxation: 
\begin{equation}
    E_{\mathrm{trap}}=E_{\mathrm{initial}}^{q}-E_{\mathrm{final}}^{q}
\end{equation}
Where $E_{\mathrm{initial}}^{q}$ and $E_{\mathrm{final}}^{q}$ are the energies of the charged cell before and after geometry optimisation respectively. A positive value indicates a more stable final state relative to the initial state.

Trap depths were reported relative to the conduction band minimum of the relaxed charged cells. A consistent eigenvalue-based definition is:
\begin{equation}
    E_{\mathrm{depth}} = E_{\mathrm{CBM}}^{q} - \varepsilon_{\mathrm{trap}}^{q},
    \label{eq:trap_depth}
\end{equation}
where $E_{\mathrm{CBM}}^{q}$ is the conduction band minimum of the relaxed charged HSE06-optimised cell and $\varepsilon_{\mathrm{trap}}^{q}$ is the occupied trap-state eigenvalue. Positive values then denote occupied trap states below the conduction band edge.

\subsection{Trap identification and electronic localisation}
\label{subsec:trap_identification_localisation_methods}

Electron localisation was characterised by orbital visualisation, Mulliken charge analysis, and the inverse participation ratio (IPR). For a molecular orbital $\Psi_k(\mathbf{r})$ expanded in an orthonormal basis $\phi_i(\mathbf{r})$ with $M$ basis functions,
\begin{equation}
    \Psi_k(\mathbf{r})=\sum_{i=1}^{M} c_{ki}\phi_i(\mathbf{r}),
    \label{eq:mo_expansion}
\end{equation}
with
\begin{equation}
    \mathrm{IPR}(\Psi_k)=
    \frac{\sum_{i=1}^{M} c_{ki}^{4}}
    {\left(\sum_{i=1}^{M} c_{ki}^{2}\right)^2}.
    \label{eq:ipr}
\end{equation}
For a state uniformly delocalised over all $M$ basis functions, $\mathrm{IPR}=1/M$; the value increases with localisation and reaches 1 for localisation on a single basis function.

A cell was classified as non-trapping (NT) when the injected electron occupied a conduction band-like state rather than a localised gap state. Localised one-electron traps were classified by the relaxed geometry of the trap. Edge traps (Ed) correspond to localisation on a widened O--Si--O edge of a SiO$_4$ unit. Face traps (F) correspond to localisation on a face of a SiO$_4$ unit. Ed and F traps were separated using the largest relaxed O--Si--O angle, with Ed traps satisfying O--Si--O$_{\max}\geq 143^\circ$. This differentiation applies to both 1- and 2- electron Ed and F traps.

Si$_3$ traps were identified when trapping was accompanied by rupture of one Si--O bond at the trapping angle. A critical bond length of approximately $BL_c\simeq 2.03$ \AA{} defines the transition from Ed-like strain to Si--O bond rupture. This differentiation applies to both 1- and 2- electron Ed and Si$_3$ traps.

Two-electron traps were classified as bi-Ed, bi-Si$_3$, or Si--Si-bonded combinations of Ed, F, and Si$_3$ motifs.

IETs containing an Si-Si trap were identified using the trap depth distributions listed in \tabref{tab:trap_taxonomy}, and labelled using an en dash between their two constituent motifs (e.g. F--F for a one-electron trap with two Si--Si-bonded F motifs). Si--Si bonded traps are distinctly deeper than their isolated counterparts: the F--F IET has a depth of 4.0~eV compared to 3.1-3.4~eV for isolated F IETs; Si--Si-bonded bipolarons containing an Si$_3$ motif have depths of 5.4-6.5~eV compared to 3.0-4.4~eV for isolated bi-Si$_3$ IETs; Si--Si-bonded bipolarons containing only Ed and F motifs have depths of 4.5-5.8~eV compared to 2.8-3.5~eV for isolated bi-Ed IETs.

\subsection{Precursor-site scoring}
\label{subsec:precursor_score_methods}

A precursor site was defined as an O--Si--O angle in the neutral defect-free model on which an injected electron localises after relaxation of the $q=-1$ cell. The precursor analysis focused on Ed traps, because the number of F traps was too small to establish robust formation criteria. Thus, only those models which trapped the electron on a site opening an O--Si--O angle to at least 143$^{\circ}$ were used to build the precursor score's definition. 

For each candidate site $i$, a set of $N$ structural descriptors $x_{ij}$ was calculated from the neutral geometry. The score of site i is defined as:
\begin{equation}
    s_i=\sum_j w_j z_{ij}, 
    \qquad
    z_{ij}=(x_{ij}-\mu_j)/\sigma_j,
    \label{eq:precursor_score_current}
\end{equation}
where $\mu_j$ and $\sigma_j$ are respectively the mean and standard deviation of descriptor $j$ over the 200-model representative subset, and $\omega_j$ is a corresponding weight between 0 and 1. The score is constructed so that larger values indicate sites more favourable for electron trapping, and the weights were optimised using a grid search.

The precursor score was then used to rank sites within each model. The site with the largest score was assigned a rank of 1 and the site with lowest score had a rank of 384. A rank-1 prediction occurs when the actual precursor site has the largest score in its cell. Top-$k$ performance was measured by the fraction of actual precursor sites that appeared among the $k$ highest-scoring sites in their respective cells.

Thirty-six candidate descriptors of short and medium range order were considered for the precursor score. The best performing score, i.e the score that assigned a rank of 1 to the most actual precursor sites, combined the following 5 descriptors: the O--Si--O angle value, the average Si--O bond length associated with the candidate angle, the Voronoi volume of the central Si atom, the Si second-nearest-neighbour distance $d_{\mathrm{Si2nn}}$, and the maximum Si nearest-neighbour distance $d_{{\mathrm{Sinn}}}^{\mathrm{max}}$. More details on the selection of descriptors and the optimisation of weights can be found in Sec. S1 E of the SM. 

\subsection{Precursor score tail density}
\label{subsec:prec_score_tail_dens}

The average volume density ($\bar\rho_{\mathrm{tail}}$) of sites lying in the upper tail of the precursor score distribution was defined as:

\begin{equation}
    \bar\rho_{\mathrm{tail}} = \frac{1}{N}\sum_i\frac{\left|\left\{j : \exists\, k,\ s_{j,k} > h\right\}\right|}{V_i},
\end{equation}

where $s_{j,k}$ is the $k^{th}$ score of all \OSiO{} sites centred on Si atom $j$ in model $i$, $V_i$ is the volume of model $i$, N is the number of models, and $h$ is a threshold defining the upper tail of the precursor score. The value of 2.49 was assigned to $h$, as it was the smallest score for a rank-1 precursor prediction. When an electron traps on a site associated with atom j, the second injected electron is always observed to trap on the same \OSiO{} site. Additional electrons would be expected to trap on other Si atoms because of the repulsive interaction between polarons. Thus, this definition of $\bar\rho_{\mathrm{tail}}$ ensures that each j counts only once per model i, even if it has more than one $s_{j,k}$ larger than h. 

\section{Results}

The results reveal two distinct regimes of intrinsic electron trapping in amorphous silica. Single electrons predominantly form edge-centred polarons on strained O--Si--O motifs, whereas two electrons predominantly drive larger structural relaxations involving Si--O bond rupture and, in some cases, Si--Si bond formation. We first define this trap taxonomy and quantify the associated trap depths. We then use the relaxed structures to identify the structural descriptors that control trap formation in the neutral network.

The electron trapping outcomes and associated trap depths are summarised in Table~\ref{tab:trap_taxonomy}.

\begin{table}[H]
\caption{\label{tab:trap_taxonomy}
Summary of one- and two-electron intrinsic electron traps (IETs) obtained from the 200 HSE06-optimized a-SiO$_2$ models. 'Ed' and 'F' denote electrons localised on the edge and face of a \SiOfour{} tetrahedron respectively. 'Si$_3$' denotes an electron localised on a threefold-coordinated Si centre. En dashes between two trap types indicates that they are linked by a Si--Si bond. The 'bi-' prefix differentiates bipolarons from polarons. ``NT'' denotes cells in which the injected electron(s) occupied conduction band-like states rather than forming localized traps. Trap depths are given relative to the CBM. The trap depths of bi-Ed--Ed, bi-Ed--F, and bi-F--F exclude the two outliers identified in \secref{subsec:results_2elec_traps}.}
\begin{ruledtabular}
\begin{tabular}{lccc}
Injected charge & Final motif & No. of cells & Trap depth (eV) \\
\colrule
$1e^-$ & Ed & 174 & 3.1--4.3 \\
       & F & 9 & 3.1--3.4 \\
       & F--F & 1 & 4.0 \\
       & Si$_3$ & 1 & 4.8 \\
       & NT & 15 & -- \\
\colrule
$2e^-$ & bi-Si$_3$ & 136 & 3.0--4.4 \\
       & bi-Si$_3$--Ed, bi-Si$_3$--F & 13 & 5.4--6.5 \\
       & bi-Ed & 6 & 2.8--3.5 \\
       & bi-Ed--Ed, bi-Ed--F, bi-F--F & 40 & 4.5--5.8 \\
       & NT & 5 & -- \\
\end{tabular}
\end{ruledtabular}
\end{table}

\subsection{One-electron traps}
One-electron trapping is dominated by edge-centred polarons that form in a variety of medium-range environments. Of the 200 models, 185 trapped the injected electron, while 15 promoted it to a conduction band-like state. The magnitude of the total-energy stabilization upon relaxation was 0.97--2.22~eV, with an average of 1.54~eV. Among the trapped cells, 174 formed an edge trap, denoted Ed, in which the electronic density localizes on an O--Si--O edge of a SiO$_4$ unit [Fig.~\ref{fig:Fig2_1e_traps}(d)]. These Ed traps correspond to the intrinsic electron traps previously identified by El-Sayed \textit{\textit{et al.}}~\cite{El-Sayed2013IdentificationCalculations}.

The Ed geometry is a distorted trigonal bipyramid. The highest-density region of the trapped state lies near the equator of the bipyramid, centred on a widened O--Si--O trapping angle. Across the Ed population, this trapping angle relaxes to 143--179$^\circ$, with an average of 164$^\circ$. The opposing O--Si--O angle relaxes to 94--126$^\circ$, with an average of 104$^\circ$. The two Si--O bonds forming the trapping angle elongate from an average initial length of 1.64~\AA{} to 1.80~\AA{}, while the two Si--O bonds forming the opposing angle elongate from an average 1.62~\AA{} to 1.70~\AA{}. Mulliken analysis shows an average decrease of 0.25~$|e|$ in the partial charge of the central Si atom. By comparison, the partial charges of the O atoms on both the trapping and opposing angle decrease only by an average 0.01~$|e|$. However, the O $p$-like molecular orbitals on and around the trap become more localized [Fig.~\ref{fig:Fig2_1e_traps}(g)]. These orbitals form the localized valence-derived band highlighted in Fig.~\ref{fig:Fig2_1e_traps}(a).

The Ed trap depth is controlled primarily by angular relaxation. The occupied Ed levels lie 3.1--4.3~eV below the CBM. As shown in Fig.~\ref{fig:Fig2_1e_traps}(b), deeper traps occur when the relaxed trapping angle approaches 180$^\circ$ and the opposing angle approaches 120$^\circ$, \textit{i.e.}, when the local structure approaches the ideal trigonal-bipyramidal geometry. The relaxed Si--O bond lengths vary over a narrower range, 1.75--1.87~\AA{}, and show a weaker apparent correlation with trap depth.

The remaining one-electron traps form a small set of minority motifs. As the trapping angle narrows, the maximum of the electronic density shifts from an edge toward a face of the SiO$_4$ unit [Fig.~\ref{fig:Fig2_1e_traps}(e)]. In the limiting case, three O atoms occupy the equatorial region of the trigonal bipyramid and one O atom lies on a pole; we denote this motif F [Fig.~\ref{fig:Fig2_1e_traps}(f)]. Only nine F traps were found, so their structural distribution is not statistically converged. The Ed--F distinction is therefore operational. The largest relaxed O--Si--O angle, \OSiOmax{}, separates most Ed and F traps [Fig.~\ref{fig:Fig2_1e_traps}(c)], although two F-like outliers lie in the apparent gap between 130$^\circ$ and 143$^\circ$. We therefore use $O$--$Si$--$O_{\max}=143^\circ$ as the cutoff between Ed and F traps in this dataset.

The F traps are shallow, with depths of 3.1--3.4~eV below the CBM. Two one-electron outliers are substantially deeper. One is an F--F motif, in which two face-like sites are connected by a Si--Si bond; this state lies 4.0~eV below the CBM [Fig.~\ref{fig:Fig2_1e_traps}(h)]. The other contains a broken Si--O bond and a threefold-coordinated Si centre, denoted Si$_3$; this state lies 4.8~eV below the CBM [Fig.~\ref{fig:Fig2_1e_traps}(i)]. These Si--Si-linked and Si$_3$ one-electron traps are rare, indicating that they require a favourable arrangement of neighbouring SiO$_4$ units. As shown in \secref{subsec:results_2elec_traps}, the same motifs become far more common after injection of a second electron, where stronger electron--lattice coupling drives larger structural relaxation.

The \OSiO{} $> 132^\circ$ precursor rule does not successfully apply to Ed traps in our results. The \OSiOmax{} of Ed precursor sites range from $111^\circ$ to $142^\circ$. Further, the precursor site was the largest \OSiO{} angle of the model in only 61 cases, while in 25 cases the electron trapped on a site neighbouring the largest angle. This observation highlights that while strained \OSiO{} play an important role in electron trapping, a single-descriptor precursor rule is insufficient to correctly describe trapping propensity. Instead, it appears that the polaronic behaviour of the injected electron may cause the creation of a new, more favourable, precursor site than that which existed in the pre-injection geometry. This model is further supported by the correlation between Ed trap depth and relaxed trap geometry. The electron relaxes deeper into the band gap as the site relaxes closer to the ideal trap geometry. Given the flexibility of the \aSiOtwo{} network, accounting for this behaviour requires a multi-descriptor precursor rule, which we expand on in \secref{sec:prec_sites}.

\newpage
\begin{figure}[H]
    \centering
    \includegraphics[width=0.7\linewidth]{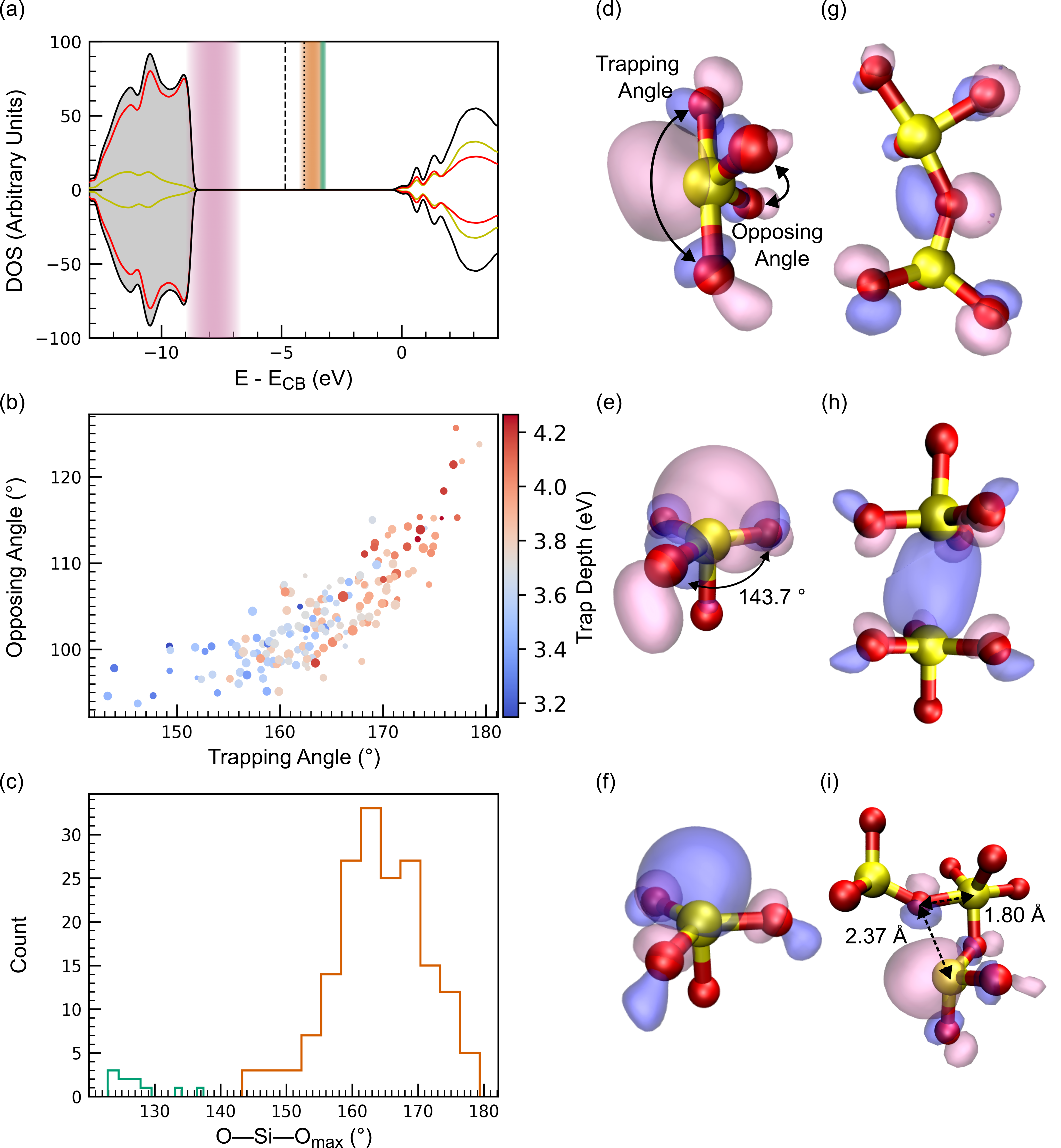}
    \newpage
    \caption{\label{fig:Fig2_1e_traps}
    One-electron intrinsic traps in a-SiO$_2$. The isosurfaces represent Kohn-Sham molecular orbitals and are drawn at an isovalue of $\pm$0.05 bohr$^{-3/2}$. Blue and mauve represent the negative and positive phases respectively. (a) Electronic density of states showing the Si, O, and total DOS in yellow, red, and black solid lines respectively. The grey fill indicates occupied states. The energy bands correspond to: localized O $p$-like molecular orbitals (pink), the Si$_3$ trap (dashed line), the F--F trap (dotted line), Ed traps (orange), and F traps (green). Energies are referenced to the CBM. Positive y-axis values are assigned to the spin-$\alpha$ channel and negative values to the spin-$\beta$ channel.(b) Ed trap depth as a function of relaxed trapping and opposing angles; marker size is scaled to the average length of the two Si--O bonds forming the trapping angle. (c) Distribution of \OSiOmax{} values for Ed (orange) and F (green) traps. (d) Ed trap. (e) Ambiguous Ed/F trap. (f) F trap. (g) Localized O $p$-like molecular orbital neighboring a trap. (h) F--F trap. (i) Si$_3$ trap.}
\end{figure}

\subsection{Two-electron traps}
\label{subsec:results_2elec_traps}
Adding a second electron changes the dominant trapping mechanism from structural relaxation to structural alteration via bond breaking and formation. In 195 of the 200 cells, the two injected electrons occupied the same molecular orbital, forming a bipolaronic trap. The magnitude of the relaxation stabilization was 0.51--4.24~eV, with an average of 1.35~eV. In the remaining five cells, both electrons occupied conduction band-like states. No cell trapped only one of the two electrons.

The dominant bipolaron is the broken-bond bi-Si$_3$ motif. In this configuration, one Si--O bond ruptures and the O atom moves away from the trap, leaving a threefold-coordinated Si centre. Isolated bi-Si$_3$ traps formed in 136 cells. A further 13 cells formed Si$_3$-containing Si--Si paired traps, so that 149 of the 195 trapped two-electron cells contained an Si$_3$ component. In 94 of these Si$_3$-containing cells, the displaced O atom remained undercoordinated, denoted O$_1$. In the remaining 55, the displaced O atom bonded to a new Si atom, forming an O$_{\mathrm{Si5}}$ configuration [Fig.~\ref{fig:Figure3_2e_trapping}(c)].

The isolated bi-Si$_3$ levels lie 3.0--4.4~eV below the CBM. Their depth increases with the separation between the Si$_3$ centre and the displaced O atom [Fig.~\ref{fig:Figure3_2e_trapping}(b)]. The deepest isolated bi-Si$_3$ traps occur when the O atom has migrated farthest from the Si$_3$ site and formed an O$_{\mathrm{Si5}}$ configuration. As in the one-electron case, bipolaron formation pulls localized O $p$-like molecular orbitals out of the valence band [Fig.~\ref{fig:Figure3_2e_trapping}(a)]. This valence-derived localized band is broader than for one-electron traps because the two-electron relaxation involves the displaced O atom and a larger distortion of the neighbouring network.

Isolated Ed and F bipolarons are much less stable than their one-electron analogues. Only six isolated bi-Ed traps remained after relaxation, and no isolated bi-F traps were observed. The isolated bi-Ed levels are the shallowest bipolarons in the dataset, lying 2.8--3.5~eV below the CBM [Fig.~\ref{fig:Figure3_2e_trapping}(d)]. Edge and face bipolarons can, however, be stabilized by forming a Si--Si bond with a neighbouring Ed, F, or Si$_3$ component. The resulting Si--Si distances are 2.26--2.65~\AA{}. We label these paired traps by their two components; for example, bi-Ed--F denotes a Si--Si-bonded pair formed from Ed and F components.

All Ed/F/Si$_3$ pair combinations were observed except for double-Si$_3$ pairs. Si--Si pairs containing an Si$_3$ component are the deepest two-electron traps, with levels 5.4--6.5~eV below the CBM [Fig.~\ref{fig:Figure3_2e_trapping}(e)]. Ed/F-only pairs lie 4.5--5.8~eV below the CBM, except for two shallower outliers in which the two components share an O atom: one bi-Ed--Ed trap at 4.0~eV and one bi-Ed--F trap at 3.9~eV below the CBM [Fig.~\ref{fig:Figure3_2e_trapping}(f)]. 

These results demonstrate that IET behaviour cannot be inferred from polaron simulations alone. The trapping of a second electron follows a fundamentally different relaxation pathway, with 97\% of cases causing an alteration of the amorphous network. Identifying newly created Si--Si bonds is straightforward using trap depth distributions: Si--Si bonded traps are distinctly deeper than their isolated counterparts. By contrast, the trap depth distributions of bi-Ed and bi-Si$_3$ motifs overlap, hence the identification of broken \SiO{} bonds requires a more careful structural analysis. 

\newpage
\begin{figure}[H]
    \centering
    \includegraphics[width=\linewidth]{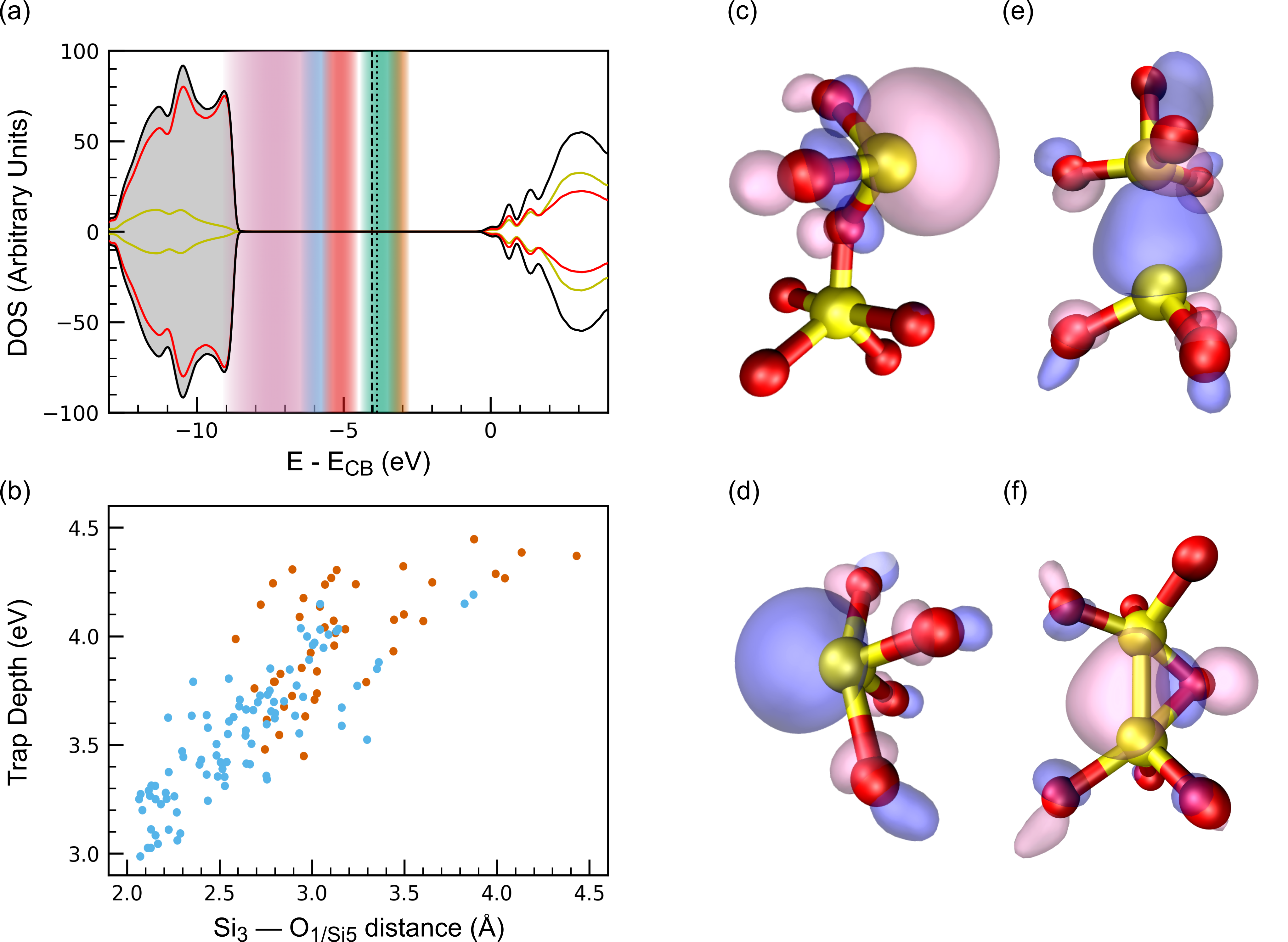}
    \caption{\label{fig:Figure3_2e_trapping}
    Two-electron intrinsic traps in a-SiO$_2$. The isosurfaces represent Kohn-Sham molecular orbitals and are drawn at an isovalue of $\pm$0.05 bohr$^{-3/2}$. Blue and mauve represent the negative and positive phases respectively. (a) Electronic density of states showing the Si, O and total DOS in yellow, red and black solid lines respectively. The grey fill indicates occupied states. The energy bands correspond to: localized O $p$-like molecular orbitals (pink), Si$_3$-containing Si--Si paired traps (blue), Ed/F-only Si--Si paired traps (red), the bi-Ed–Ed outlier (dashed line), the bi-Ed–F outlier (dotted line), isolated bi-Si$_3$ traps (green), and isolated bi-Ed traps (orange). Energies are referenced to the CBM. Positive y-axis values are assigned to the spin-$\alpha$ channel and negative values to the spin-$\beta$ channel. (b) Isolated bi-Si$_3$ trap depth as a function of the distance between the Si$_3$ centre and the displaced O atom. Blue and orange markers identify O$_1$ and O$_{\mathrm{Si5}}$ configurations respectively. (c) bi-Si$_3$ trap neighbouring an O$_{\mathrm{Si5}}$ site. (d) bi-Ed trap. (e) bi-Si$_3$--F trap. (f) Shallow bi-Ed--Ed outlier in which the two trap components share an O atom.}
\end{figure}

\newpage
\subsection{Boundary between Ed and Si$_3$ bipolaron motifs}
\label{subsec:Ed_vs_Si3}

The Ed \textit{vs.} F distinction is less important for two-electron traps than for one-electron traps because isolated bi-Ed and bi-F traps are rare, and because Ed/F-containing Si--Si pairs have similar trap depths. We therefore retain the same operational cutoff, \OSiOmax{}$=143^\circ$, to distinguish Ed and F components.

The Ed \textit{vs.} Si$_3$ distinction is more important because it marks Si--O bond rupture. We define two distances, $\mathrm{BL}_1$ and $\mathrm{BL}_2$, as shown in Fig.~\ref{fig:Figure4_E_vs_Si3}(a). For Ed motifs, $\mathrm{BL}_1$ and $\mathrm{BL}_2$ are the longer and shorter Si--O bonds forming the trapping edge. For Si$_3$ motifs, $\mathrm{BL}_1$ is the distance between the displaced O atom and the Si$_3$ centre, while $\mathrm{BL}_2$ is the remaining Si--O bond associated with the original trapping edge. We also define the angle $\gamma$ between $\mathrm{BL}_1$ and $\mathrm{BL}_2$ such that:
\begin{equation}
    \gamma = \begin{cases}
                \cos^{-1}\left(\hat{\mathbf{v}}_1 \cdot \hat{\mathbf{v}}_2\right) & \text{if } (\hat{\mathbf{v}}_1+\hat{\mathbf{v}}_2) \cdot (\hat{\mathbf{v}}_3+\hat{\mathbf{v}}_4) \leq 0, \\
                360 - \cos^{-1}\left(\hat{\mathbf{v}}_1 \cdot \hat{\mathbf{v}}_2\right) & \text{if } (\hat{\mathbf{v}}_1+\hat{\mathbf{v}}_2) \cdot (\hat{\mathbf{v}}_3+\hat{\mathbf{v}}_4) > 0,
             \end{cases}
    \label{eq:gamma}
\end{equation}
with
\begin{equation}
    \hat{\mathbf{v}}_i =
    \frac{\mathbf{r}_{\mathrm{O}_i}-\mathbf{r}_{\mathrm{Si}_0}}
    {\left|\mathbf{r}_{\mathrm{O}_i}-\mathbf{r}_{\mathrm{Si}_0}\right|},
    \quad i \in \{1,2,3,4\}.
    \label{eq:tgamma_vectors}
\end{equation}

Figure~\ref{fig:Figure4_E_vs_Si3}(b) shows two regimes. For short $\mathrm{BL}_1$, both $\mathrm{BL}_1$ and $\mathrm{BL}_2$ elongate, consistent with symmetric strain of the trapping edge. This regime persists up to a critical length $\mathrm{BL}_c \simeq 2.03$~\AA{}. Beyond $\mathrm{BL}_c$, $\mathrm{BL}_1$ continues to increase while $\mathrm{BL}_2$ relaxes toward approximately 1.8~\AA{}, signalling the loss of the original Si--O bond. The angular coordinate shows the same crossover: $\gamma$ increases from the Ed range to values above 200$^\circ$, remains between 180$^\circ$ and 220$^\circ$ during the onset of bond rupture, and becomes widely distributed once the Si--O bond is fully broken.

Mulliken charges provide an independent marker of this structural transition. The partial charge of the displaced O atom initially changes with increasing $\mathrm{BL}_1$, but the trend begins to plateau beyond $\mathrm{BL}_c$ [Fig.~\ref{fig:Figure4_E_vs_Si3}(c)]. We therefore use $\mathrm{BL}_c \simeq 2.03$~\AA{} as the operational boundary between Ed-like and Si$_3$-like bipolaron components. Si--Si-containing motifs cluster on the relaxed side of this boundary, indicating that Si--Si bonding stabilises configurations in which the original Si--O bond has ruptured.
\newpage
\begin{figure}[H]
    \centering
    \includegraphics[width=\linewidth]{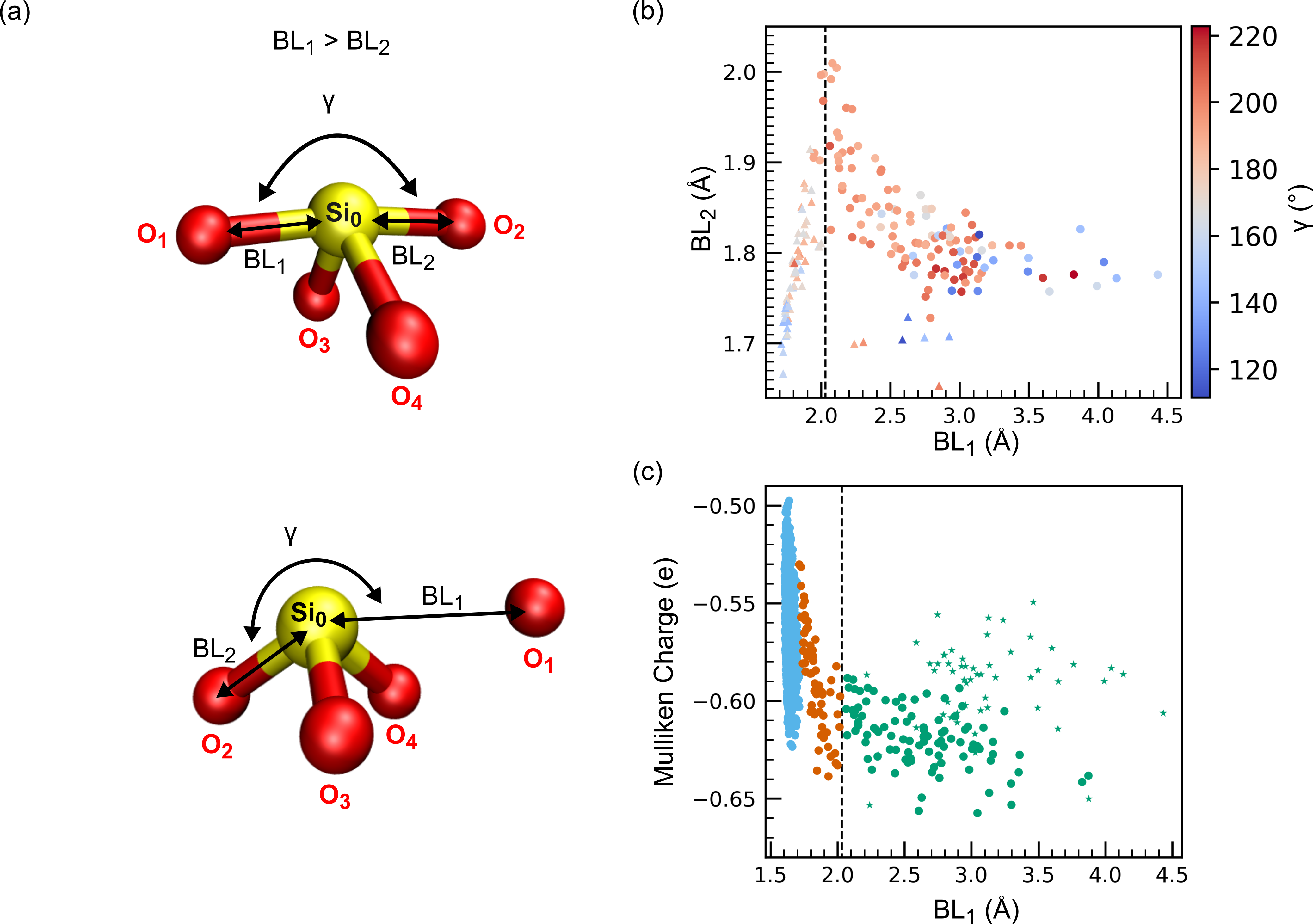}
    \caption{\label{fig:Figure4_E_vs_Si3}
    Structural boundary between Ed and Si$_3$ bipolaron components. (a) Definition of $\mathrm{BL}_1$, $\mathrm{BL}_2$, and $\gamma$ for Ed and Si$_3$ sites. (b) Relationship between $\mathrm{BL}_1$ and $\mathrm{BL}_2$, showing the crossover from symmetric edge strain to Si--O bond rupture near $\mathrm{BL}_c \simeq 2.03$~\AA{}. Triangles correspond to trap sites that are part of an Si--Si pair, while circles correspond to isolated trap sites. (c) Mulliken partial charge of O as a function of $\mathrm{BL}_1$ for: O bonded to a non-defective Si centre (blue), O on trapping angles (orange), O$_1$ (green circles), and O$_\mathrm{Si5}$ (green stars). For non-defective, fourfold-coordinated Si centres, $\mathrm{BL}_1$ is defined as its longest \SiO{} bond. Each blue point therefore corresponds to the O forming the longest \SiO{} bond for each non-defective Si centre.}
\end{figure}

\subsection{Bipolaron-induced network distortion}

The formation of bipolarons distorts the surrounding amorphous network. Figure~\ref{fig:Figure5_network_weakening}(a) plots the bond-angle and bond-length distortion of SiO$_4$ units as a function of their distance from the nearest bipolaron trap. Most tetrahedra remain weakly distorted and lie close to the origin, as illustrated in Fig.~\ref{fig:Figure5_network_weakening}(b). Strong distortions are concentrated within approximately 5~\AA{} of the trap. In these neighbouring units, two O--Si--O angles often open simultaneously, flattening the tetrahedron and elongating the associated Si--O bonds [Fig.~\ref{fig:Figure5_network_weakening}(c)].

This distortion shows that trap formation is not controlled solely by the local O--Si--O angle of the eventual trap site. The flexibility of the surrounding network also determines whether the local SiO$_4$ unit can accommodate the angular opening and bond elongation needed to stabilise the trapped charge. This observation motivates the precursor-site analysis below, which incorporates both local strain and medium-range network flexibility.

\begin{figure}[H]
    \centering
    \includegraphics[width=\linewidth]{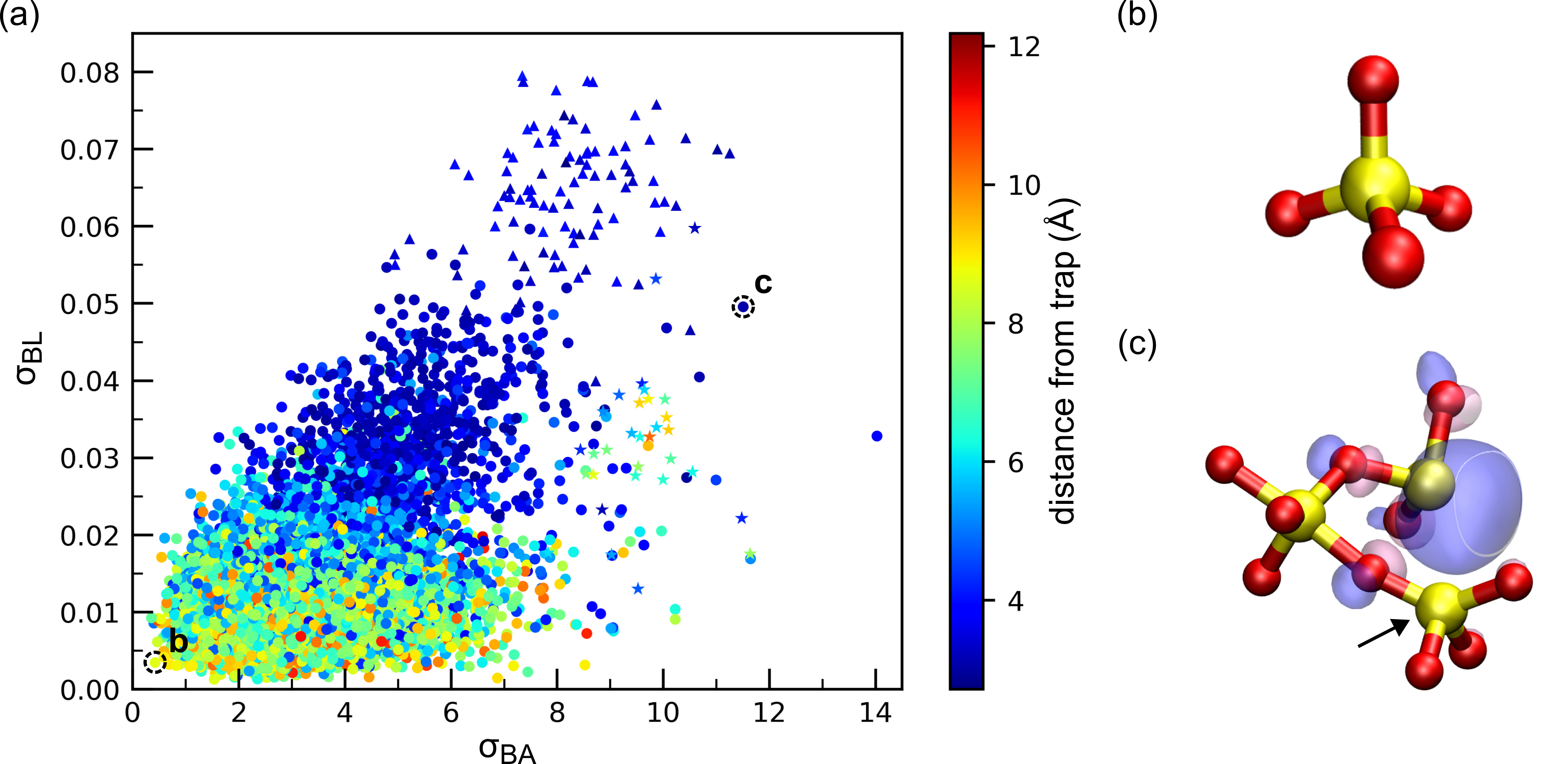}
    \caption{\label{fig:Figure5_network_weakening}
    Bipolaron-induced distortion of neighbouring SiO$_4$ units. (a) Bond-angle and bond-length distortion of \SiOfour{} units mapped against the distance from the nearest bipolaron trap. Stars denote \SiOfour{} units that are part of two-membered rings, triangles denote \SiOfour{} units that are bonded to a onefold-coordinated O, and circles are used for all other \SiOfour{} units. The highlighted points b and c correspond to the units shown in sub-figures (b) and (c) respectively. (b) Weakly distorted SiO$_4$ unit. (c) Strongly distorted SiO$_4$ unit neighbouring a bipolaron. The arrow points to the strongly distorted unit, which corresponds to point c in (a).}
\end{figure}

\subsection{Precursor sites}
\label{sec:prec_sites}

We identify precursor sites by correlating the neutral-cell geometry with the relaxed one-electron trapping outcome. We focus on Ed traps because they form the only statistically robust one-electron trap population in the present dataset. An Ed precursor is defined as the neutral-cell O--Si--O angle that becomes the trapping angle after electron injection and structural relaxation.

An O--Si--O angle criterion alone does not identify most precursor sites. The neutral O--Si--O angles that later formed Ed traps ranged from 111$^\circ$ to 142$^\circ$, with an average of 120$^\circ$. The O--Si--O angle nevertheless remains an important descriptor: in 61 cells, the trapping angle was the largest O--Si--O angle in the neutral structure. In a further 21 cells, the electron trapped on an angle neighbouring the largest O--Si--O angle (\textit{i.e.} having at least one atom in common with the largest \OSiO{}). Likewise for bond lengths: the maximum bond length of precursor sites varied between 1.59 and 1.68 \AA{}, 30 electrons trapped on a site with the longest bond in the model, and 44 on a neighbouring site (having one atom in common). While strained bonds and \OSiO{} angles clearly facilitate trapping, neither are a necessary nor sufficient condition for electron trapping because the injected electron can drive a cooperative local relaxation, where its interaction with the atomic network can create a new local geometry that is more favourable for self-trapping.

The neutral electronic structure is further insufficient to reliably predict precursor sites. Figure~\ref{fig:Figure6_IPRs} compares two representative cells using the inverse participation ratio (IPR). In one case, the lowest unoccupied molecular orbital already shows partial localization before electron injection [Fig.~\ref{fig:Figure6_IPRs}(a)], and relaxation further localizes the occupied state into the band gap [Fig.~\ref{fig:Figure6_IPRs}(b)]. In the second case, the LUMO is initially delocalized [Fig.~\ref{fig:Figure6_IPRs}(c)], but the injected electron still forms a localized polaron after relaxation [Fig.~\ref{fig:Figure6_IPRs}(d)]. Precursor identification must therefore capture not only pre-existing electronic localization, but also the capacity of the network to self-trap the electron through structural relaxation.

\begin{figure}[H]
    \centering
    \includegraphics[width=\linewidth]{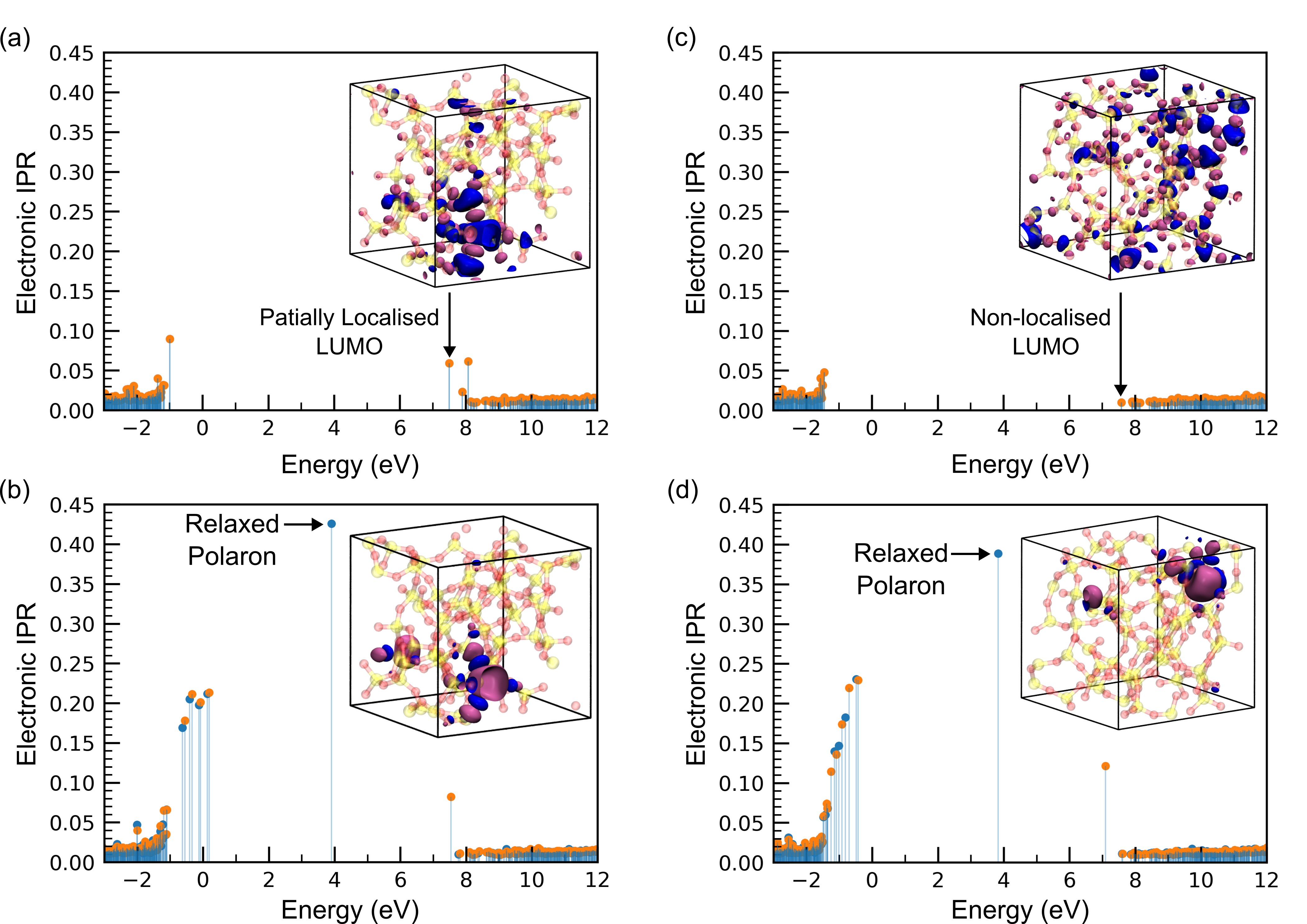}
    \caption{\label{fig:Figure6_IPRs}
    Electronic localization before and after one-electron trapping. The IPRs of the molecular orbitals are plotted using blue dots for the alpha spin channel and orange dots for the beta spin channel. The isosurfaces represent Kohn-Sham molecular orbitals and are drawn at a value of $\pm$0.02 bohr$^{-3/2}$. Blue and mauve isosurfaces represent the negative and positive phases respectively. The atomic structures are shown with low opacity for visual clarity. (a),(b) Cell in which the neutral LUMO is already partially localized before electron injection and becomes more localized after polaron relaxation. (c),(d) Cell in which the neutral LUMO is initially delocalized but forms a localized polaronic state after relaxation.}
\end{figure}

Medium-range order provides the missing information. We characterize the medium-range environment of each O--Si--O angle by the $d_{\mathrm{Si2nn}}$ distance and the $\xi$ angle defined in the Methods. This packing distribution is shown in Fig.~\ref{fig:Figure7_MRO_packing}(a). The isolated island in the lower-left region corresponds to three-membered rings [Fig.~\ref{fig:Figure7_MRO_packing}(d)], which have a narrow range of $d_{\mathrm{Si2nn}}$. The lowest-distance outliers correspond to three-membered rings that are adjacent to a two-membered ring.

The packing distribution links angular strain to the geometry of neighbouring tetrahedra. The widest O--Si--O angles, above 130$^\circ$, occur primarily at $d_{\mathrm{Si2nn}}$ above approximately 5.5~\AA{} [Fig.~\ref{fig:Figure7_MRO_packing}(c)]. Other strained angles, above 120$^\circ$, arise from less direct combinations of $d_{\mathrm{Si2nn}}$ and $\xi$ angle [Fig.~\ref{fig:Figure7_MRO_packing}(b)]. The Ed precursors cluster toward large $d_{\mathrm{Si2nn}}$ and large $\xi$ angles. Sites in this region are flexible because widening the trapping O--Si--O angle can be accommodated by rotations of neighbouring Si--O--Si linkages, while the adjacent Si atoms are sufficiently separated to allow Si--O bond elongation. By contrast, sites near the centre of the distribution [Fig.~\ref{fig:Figure7_MRO_packing}(e)] require larger collective displacements or place additional strain on neighbouring tetrahedra, making electron trapping less favourable.

\begin{figure}[htpb]
    \centering
    \includegraphics[width=\linewidth]{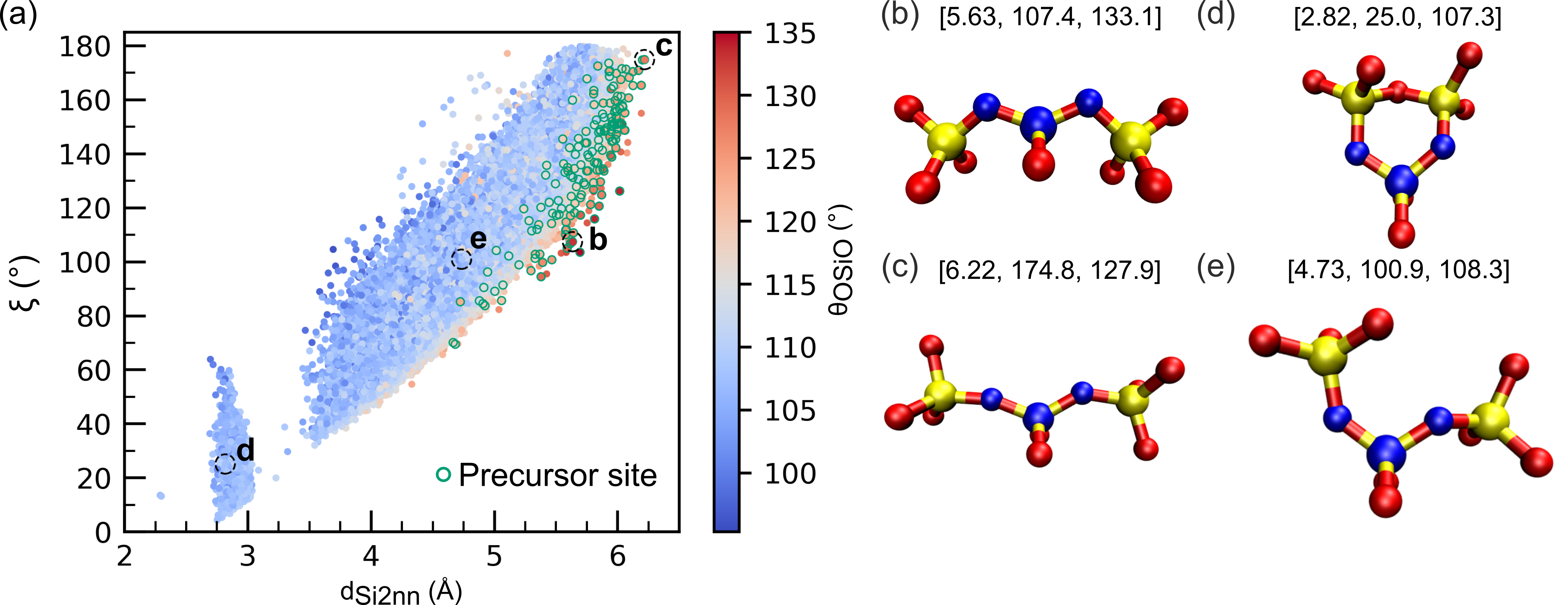}
    \caption{\label{fig:Figure7_MRO_packing}
    Medium-range packing environment of O--Si--O angles in neutral a-SiO$_2$. (a) Packing distribution using the $d_{\mathrm{Si2nn}}$ distance and $\xi$ angle. Colour indicates the neutral O--Si--O angle; green outlines indicate Ed precursor sites. The highlighted points b, c, d, and e correspond to the structures shown in sub-figures (b), (c), (d), and (e) respectively. The square brackets over each sub-figure contains the coordinates of the corresponding highlighted point: [$d_{\mathrm{Si2nn}}$, $\xi$, $\theta_{\mathrm{OSiO}}$] (b) Strained site with low $\xi$ angle. (c) Strained site with large $d_{\mathrm{Si2nn}}$ and large $\xi$ angle. (d) Three-membered ring. (e) Representative less-flexible site from the central region of the distribution.}
\end{figure}

Because no single descriptor captures all precursor sites, we rank candidate O--Si--O angles using the composite precursor score defined in the Methods. The descriptor set is listed in Table~\ref{tab:score_params}, their distributions are given in Fig. S6 of the SM. The selected descriptors separate three contributions: initial local strain, captured by the O--Si--O angle and average Si--O bond length; local free volume, captured by the Si Voronoi volume and maximum Si nearest-neighbour distance; and medium-range flexibility, captured by the Si second-nearest-neighbour distance. Since the same dataset was used to select the descriptor set and evaluate the ranking, we interpret the score as a descriptive measure of trapping propensity rather than an independently validated predictive model.

\begin{table}[H]
\caption{\label{tab:score_params}
Descriptors included in the composite precursor score. The weights are those used in the ranking analysis; descriptor centring is defined in the Methods.}
\begin{ruledtabular}
\begin{tabular}{lcc}
Descriptor & Structural role & Weight \\
\colrule
$\theta_{\mathrm{OSiO}}$ : O--Si--O angle & Local angular strain & 0.7 \\
$d_{\mathrm{Si-O}}^{\mathrm{avrg}}$ : Average Si--O bond length & Local bond strain & 0.3 \\
$V_{\mathrm{Si_0}}$ : Si Voronoi volume & Local free volume & 0.2 \\
$d_{\mathrm{Si2nn}}$ : Si second-nearest-neighbour distance & Medium-range strain & 0.5 \\
$d_{\mathrm{Sinn}}^{\mathrm{max}}$ : Maximum Si nearest-neighbour distance & Local network openness & 0.4 \\
\end{tabular}
\end{ruledtabular}
\end{table}

Each 192-atom \aSiOtwo{} cell contains 384 candidate O--Si--O angles. Table~\ref{tab:score_rankings} compares the cumulative top-$k$ recall obtained from the O--Si--O angle alone with that obtained from the composite score. Using the O--Si--O angle alone, 35\% of Ed precursors were ranked first in their cell and 65\% were ranked within the top five. The composite score improves these values to 60\% and 81\%, respectively. All Ed precursors are recovered within the top 40 candidates using the composite score, whereas the O--Si--O angle alone requires the top 113 candidates to recover all precursors. Our dataset thus shows that earlier single-geometry criteria capture a subset of highly strained sites, but miss the majority of active precursors because they do not describe the relaxation pathway available to the surrounding network.

\begin{table}[H]
\caption{\label{tab:score_rankings}
Comparison of the cumulative top-$k$ recall of Ed precursor sites when using a precursor score based on the \OSiO{} angle only \textit{vs.} when using the composite score defined in \tabref{tab:score_params}. Entries of the second and third columns give the percentage of Ed precursors whose neutral-cell O--Si--O angle is ranked within the top $k$ candidates of the corresponding cell.}
\begin{ruledtabular}
\begin{tabular}{ccc}
Rank $k$ & O--Si--O angle only (\%) & Composite score (\%) \\
\colrule
1   & 35  & 60  \\
2   & 47  & 69  \\
3   & 57  & 74  \\
4   & 62  & 78  \\
5   & 65  & 81  \\
40  & 94  & 100 \\
113 & 100 & 100 \\
\end{tabular}
\end{ruledtabular}
\end{table}

Non-selected sites are not necessarily unable to trap an electron; rather, they were not selected in the unconstrained relaxation. This interpretation is supported by additional tests in which freezing the atoms of the original precursor allowed the electron to trap at an alternative site. Out of 6 cells used in this test, 2 did not trap the electron again, 2 trapped it on a site neighbouring the original trap site and 2 trapped it on another non-neighbouring site. In particular, non-selected sites that were ranked higher than the precursor sites were found to have less local strain: an average \OSiO{} angle of 116~$^\circ$ against 120~$^\circ$ for precursor sites, and an average \SiO{} bond length of 1.63~\r{A} against 1.64~\r{A} for precursor sites. This highlights that trap formation is favoured under two conditions: a local geometry that is already strained, and a surrounding network flexible enough to accommodate the polaronic relaxation.

\section{Discussion}

    \subsection{Precursor site properties}
        Other precursor site definitions have been put forward in previous studies. El-Sayed \textit{et al.} defined them as an O--Si--O angle of at least 132$^\circ$ \cite{El-Sayed2013IdentificationCalculations} while Li \textit{et al.} suggested that an IET precursor containing both a wide O--Si--O angle ($\geq 132^\circ$) and a long Si--O bond ($\geq 1.660$~\AA) can undergo electron-induced Si--O bond rupture \cite{Li2023TheoreticalSilica}. Using the first definition, only 5 of the 174 polaron IETs found in our simulations would have been correctly predicted, with the smallest IET precursor \OSiO{} angle in our dataset having a value of 111$^\circ$. Likewise, we find counterexamples for Li \textit{et al.'s} definition: all the bond lengths of the single polaron Si$_3$ IET in our dataset were less than 1.64 \r{A}, and the ruptured bond was in fact the site's shortest bond at 1.625 \r{A}; Among the Si$_3$ sites in bipolaron IETs, the bond lengths of the ruptured Si-O bonds prior to the injection of the second electron range from 1.59~\r{A} to 1.96~\r{A} with an average of 1.81 \r{A}. The disagreement with earlier precursor rules reflects the need for statistical significance to robustly characterise IETs and establish transferable precursor rules. Small melt-quench model ensembles may contain valid examples of trapping motifs but  cannot capture a representative distribution of the critical medium-range environments which determine whether the site can relax into a stable trap configuration.

        Furthermore, we find a similar concentration of O--Si--O angles exceeding 132$^\circ$ ($9.1 \times 10^{18} / cm^3$) as in the study by El-Sayed \textit{et al.} ($4.4 \times 10^{19} / cm^3$). However, we cannot exclude the possibility that an O--Si--O angle larger than 132 degrees may ultimately be unable to trap the electron because of a constraining medium-range environment. We expect that the minimum O--Si--O angle that would guarantee to trap an electron should be around 144$^\circ$, \textit{i.e.} the smallest trapping angles observed in our simulations. Such angles should trap the electron with minimal structural relaxation. 
        
        It is not possible to give a meaningful estimate of the concentration from our results because we observe that cells may contain more than one precursor site. We also expect that the concentration of trapped bipolarons may be affected by a repulsive interaction between them, a topic which requires further study. Nevertheless, it is important to have at least a qualitative understanding of the impact of choices such as the quench rate and cell size on the number of precursor sites. The precursor score tail density $\bar\rho_{\mathrm{tail}}$ was developed for this purpose and is defined in the Methods. Figure~\ref{fig:Figure9_precursor_concentration} plots the sensitivity of $\bar\rho_{\mathrm{tail}}$ to the quench rate for 192-atom and 648-atom cells. The calculation of $\bar\rho_{\mathrm{tail}}$ was performed immediately after the MD simulations. A comparison with PBE optimised models in Fig. S7(b) of the SM shows no significant change of $\bar\rho_{\mathrm{tail}}$ after PBE optimisation. The 192-atom data points are based on at least 30 cells, and the 648-atom data points are based on at least 10, sufficient for a qualitative discussion of $\bar\rho_{\mathrm{tail}}$. The size of the tail is sensitive to the quench rate and is independent of the choice of cell size. The sensitivity to the quench rate is explained by the already documented dependence of the high-strain tails of structural distributions, such as bond angle and bond length \cite{Lane2015CoolingDynamics}. This sensitivity underlies the need to explicitly test the convergence of high-strain tails of relevant property distributions to infer structural precursor rules. Furthermore, since typical quench rates in computational simulations are orders of magnitude faster than the fastest experimental quench rates, this dependence suggests that it is too early to estimate a concentration of precursor sites from melt-quench simulations. Even if the distributions of geometric properties of amorphous silica models produced with our unrealistically high quench rates remain consistent with experimental results, IET precursor sites in \aSiOtwo{} populate the outlier tails of those distributions, which remain difficult to probe experimentally.
        
        \begin{figure}[htpb]
            \centering
            \includegraphics[width=0.5\linewidth]{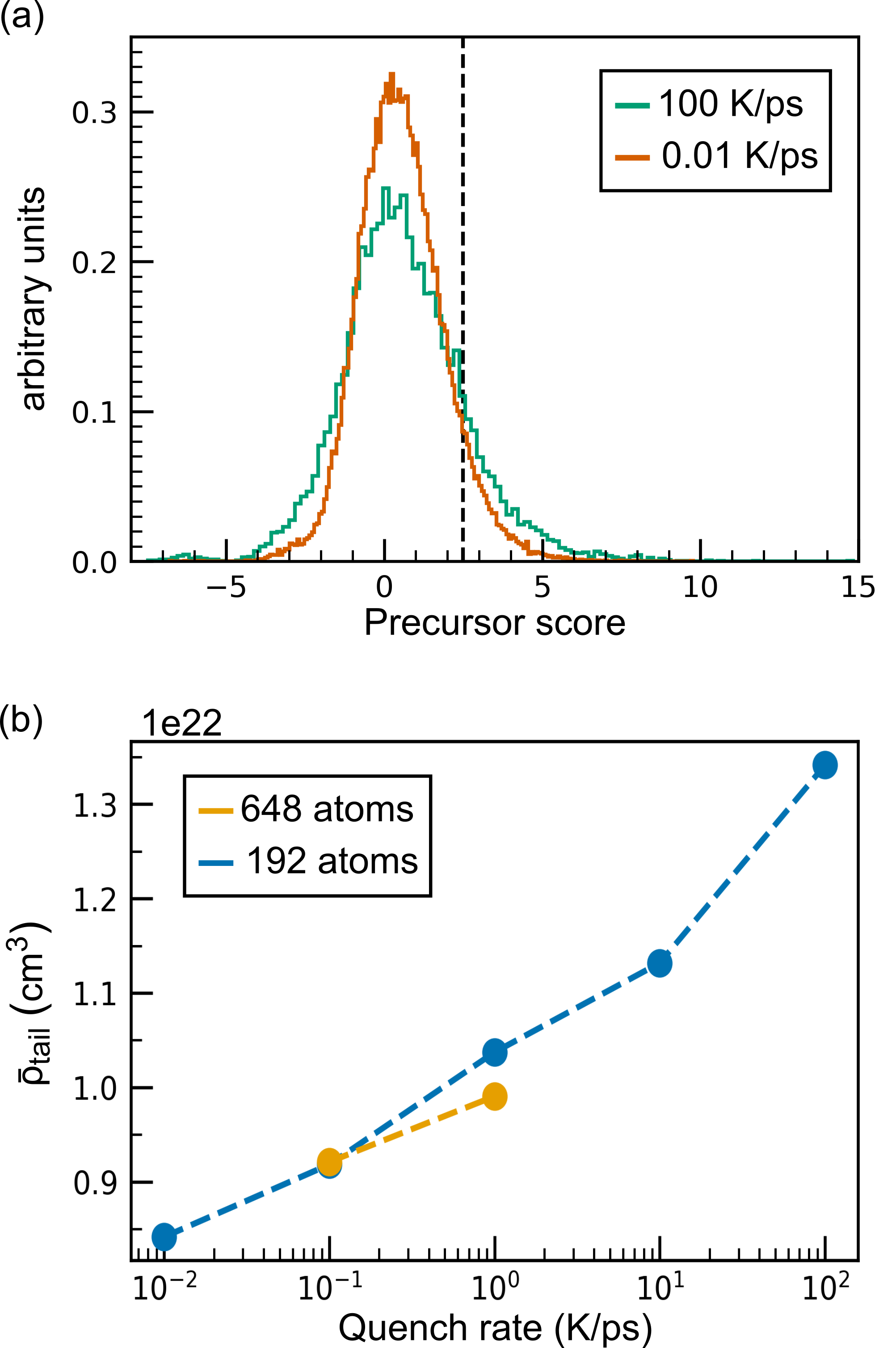}
            \caption{(a) Precursor score distributions for 192-atom cells at two different quench rates. The black dashed line identifies the threshold used to define the upper tail of the distributions (at a value of 2.49). (b) Precursor score tail density as a function of quench rate and cell size.}
            \label{fig:Figure9_precursor_concentration}
        \end{figure}

        Lastly, the choice of MD potential is expected to impact the availability of precursor sites. Li \textit{et al.} used a ReaxFF force field and reported ring size distributions centred on 5-membered rings. El-Sayed \textit{et al.} also used a ReaxFF force field and found predominantly four- and five-membered rings. The most frequent ring sizes in our models are six-, seven-, and eight-membered rings. Smaller rings have narrower O--Si--O angles, which reduces the number of strained sites in a cell, and form less flexible structures, which can hinder electron trapping. This is reflected in the frequency of electron trapping in the study by El-Sayed \textit{et al.}, where 4 out of 20 models trapped the electron. Likewise, the structure used by Li \textit{et al.} trapped the electron only once strain was applied. We observe much more frequent trapping (in 174 out of 200 models for polarons). More details on the impact of ring size on the packing distribution and electron trapping can be found in Sec. S3 B of the SM. 

    \subsection{Spin treatment of two-electron cells}
    \label{subsubsec:spin_treatment_methods}
    
    For charge state $q=-2$, both singlet and triplet configurations were considered. Their relative stability was tested on 20 models. In singlet calculations, the injected electrons either paired on the same localised site or both occupied conduction band-like states. In triplet calculations, the electrons localised on two different sites. The uncorrected energies of singlet configurations were lower than those triplet configurations for all 20 test models; the differences ranged from 0.14 eV to 2.76 eV with an average of 1.01 eV.
    
    The Lany--Zunger correction could not be directly applied to the triplet test calculations because the triplet charge density was distributed over two spatially separated sites, leading to an undefined Madelung constant. Instead, the energy difference between the corrected singlet and triplet energies were calculated using two bounding scenarios: 1) treating the charge as a single point charge and applying the average $q=-2$ finite-size correction once, or 2) treating the charge in the triplet models as two point charges and applying the average $q=-1$ finite-size correction twice. The true correction value is expected to lie between those two limits. The average $q=-2$ potential alignment correction was also applied to all triplet models.

    For case 1, all singlet configurations were once again lower in energy, with differences ranging from 0.11 eV to 3.20 eV and an average of 1.00 eV; a results comparable to the no-correction case. For case 2, four triplet configurations became lower in energy than the corresponding singlet configurations, with differences ranging from -0.43 eV to 2.65 eV and an average of 0.45 eV. Only under the most triplet-favourable correction bound do a small number of test cells become triplet-lower; these cases should be regarded as near-degenerate within the uncertainty of the charged-cell correction rather than as robust triplet ground states. The tested ensemble therefore supports the conclusion that the low-energy q = -2 response is predominantly bipolaronic. More details on the comparison of singlet and triplet $q=-2$ configurations can be found in Sec. S3 C of the SM.
    
    \subsection{Correspondence with experimental evidence}
        Assignment of optical absorption bands is controversial due to overlaps between defect types. Our results support the view than intrinsic traps at least contribute to those bands. Comparing with the optical absorption results of El-Sayed \textit{et al.} \cite{El-Sayed2015OpticalSiO2}, our band for bi-Si$_3$--Ed and bi-Si$_3$--F traps corresponds to their 6.4 eV band. Our band for bi-Ed--Ed, bi-Ed--F and bi-F--F corresponds well to their 4.7 eV band, and our band for Si$_3$ traps corresponds well to their 3.7 eV band. These results support a possible structural interpretation of the observed optical bands, although a definitive assignment would require explicit optical-transition calculations. The population statistics show that final-state trap depth does not alone determine the dominant bipolaron configuration: Si$_3$-containing Si--Si pairs are deeper than Ed/F-only Si--Si pairs, yet they are less frequent: 13 cells versus 40 cells, respectively. This contrast indicates that trap formation is also governed by the accessibility of the required structural relaxation pathway, and motivates an investigation of the energy barriers between these configurations to determine their expected relative probabilities.

\section{Conclusion}
    We present a statistically robust characterisation of the structural and electronic properties of intrinsic polaron and bipolaron intrinsic electron traps in amorphous silica. Given the close relationship between electron trapping and strain in the amorphous network, converging the high-strain structural features of \aSiOtwo{} and representative sampling were essential steps to achieve transferable precursor rules and well-defined IET populations. Electron trapping is accompanied by a structural relaxation of the trap site which localises the wavefunction of the trapped electron and leads to a state deep in the band gap. Trapping is not controlled by any single geometric descriptor, it instead arises from the combination of local strain and the ability of the surrounding amorphous network to accommodate the required polaronic or bipolaronic relaxation. Polaron trapping is predominantly mediated by structural relaxation into "Ed" configurations by widening an O--Si--O angle of the precursor site and elongating its bonds. By contrast, bipolaron trapping involves topological changes. The injection of the second electron typically causes the rupture of an Si-O bond, forming Si$_3$ trap configurations. However, the formation of an Si--Si bond can stabilise a bipolaron trap and prevent the rupture of the Si-O bond. For both polarons and bipolarons, the depth of the trap in the band gap is correlated to how well it can relax into an ideal trap geometry. The site's ability to relax is influenced by its initial strain (wide O--Si--O angle, and elongated bonds) and by the flexibility of the surrounding amorphous network (proximity and orientation of neighbouring SiO$_4$ units). In turn, sites' ability to relax determines the trap precursor site: the electron will preferentially trap at the site which will form a deep trap while overcoming a minimal energy barrier. The diversity of medium-range environments in which strained local motifs can be found explains why small ensembles of amorphous models can identify individual trapping motifs but cannot provide reliable precursor criteria or concentrations. We find that the best prediction of precursor sites is obtained by collectively considering: the value of the trapping angle, the average bond length of the trapping angle, the Voronoi volume of the central Si atom, the maximum distance of a Si nearest neighbour, and the distance between the two Si atoms on either side of the trapping angle. Finally, we associate the most common bipolaron configurations with optical absorption bands reported by \cite{El-Sayed2015OpticalSiO2} : bi-Si$_3$ IETs with the 3.7 eV band; bi-Ed--Ed, bi-Ed--F and bi-F--F IETs with the 4.7 eV band; and bi-Si$_3$--Ed and bi-Si$_3$--F with the 6.4 ev band. This study clarified the nature of intrinsic electron traps in amorphous silica, highlighted the role of the flexibility of the amorphous network in trapping electrons, and demonstrated the necessity of representative sampling for quantitative amorphous-defect modelling.

\section*{CRediT authorship contribution statement}
\noindent
\textbf{Robin Delhomme:} Conceptualization, Methodology, Software, Validation, Formal analysis, Investigation, Data curation, Visualization, Writing -- original draft, Writing -- review \& editing.
\textbf{Jonathon Cottom:} Conceptualization, Methodology, Software, Validation, Formal analysis, Investigation, Visualization, Writing -- original draft, Writing -- review \& editing.
\textbf{Emilia Olsson:} Conceptualization, Methodology, Software, Resources, Supervision, Project administration, Funding acquisition, Writing -- review \& editing.

\section*{Acknowledgements}
This work was conducted at the Advanced Research Center for Nanolithography, a public-private partnership between the University of Amsterdam (UvA), Vrije Universiteit Amsterdam (VU), Rijksuniversiteit Groningen (RUG), the Netherlands Organization for Scientific Research (NWO), and the semiconductor equipment manufacturer ASML. This work made use of the Dutch national e-infrastructure with the support of the SURF Cooperative using grant no. EINF-11683, EINF-14132, EINF-16595 and EINF-17976. The authors thank SURF (www.surf.nl) for the support in using the National Supercomputer Snellius. E.O. is grateful for a WISE Fellowship from the NWO and acknowledges support via Holland High Tech through a public-private partnership in research and development within the Dutch top sector of High-Tech Systems and Materials (HTSM).

\section*{Data availability}
The table containing the Coulomb and short-range correction terms for MD simulations is publicly available in a Zenodo repository, along with the structure files of all model ensembles \cite{zenodo_repo}.

\bibliographystyle{apsrev4-2}
\bibliography{references}

\end{document}


\title{Supplementary Information for "Statistically Robust Characterization of Polaron and Bipolaron Intrinsic Traps in Amorphous Silica Reveals the Role of Network Flexibility in Electron Localisation"}

\author{Robin Delhomme}
\affiliation{Advanced Research Center for Nanolithography, Science Park 106, 1098 XG Amsterdam, The Netherlands}
\affiliation{Institute of Theoretical Physics, Institute of Physics, University of Amsterdam, Science Park 904, 1098 XH Amsterdam, The Netherlands}

\author{Jonathon Cottom}
\affiliation{Advanced Research Center for Nanolithography, Science Park 106, 1098 XG Amsterdam, The Netherlands}
\affiliation{Institute of Theoretical Physics, Institute of Physics, University of Amsterdam, Science Park 904, 1098 XH Amsterdam, The Netherlands}

\author{Emilia Olsson}
\affiliation{Advanced Research Center for Nanolithography, Science Park 106, 1098 XG Amsterdam, The Netherlands}
\affiliation{Institute of Theoretical Physics, Institute of Physics, University of Amsterdam, Science Park 904, 1098 XH Amsterdam, The Netherlands}

\date{\today}

\maketitle

\section{Supplementary Methods Details}
    
\subsection{Generation and refinement of amorphous silica models}

\subsubsection{Generation of 648-atom models}
The procedure to generate 648-atom models was identical to the 192-atom case, except for the following: 432 O atoms and 216 Si atoms were randomly placed in box of 32 \AA{} cell length, and snapshots were generated every 40 ps.

\subsubsection{Structural decorrelation of successive snapshots}

Complete structural decorrelation was defined as the saturation of atoms' average displacement since the previous snapshot. To establish the minimum elapsed time $\Delta t$ required between snapshots, the average atomic displacement $\bar{D}$ was plotted as a function of $\Delta t$. Starting from a cell of randomly placed atoms, the cell was brought to 5000 K in 5 ps and equilibrated for 50 ps for 192-atom cells and 100 ps for 648-atom cells. After equilibration, 20 snapshots were generated spaced by $\Delta t$ ps. This procedure was repeated for values of $\Delta t$ that ranged from 5 to 50 ps in steps of 5 ps, with an additional point at 2.5 ps. The random seeds used to initialise the atom positions were kept the same in every run. The average displacements $\bar{d_j}$ of atoms in cell j, relative to the preceding cell j-1 is:

\begin{equation}
    \bar{d_j} = \frac{1}{n_{a}} \sum_{i}^{n_a}\left|\mathbf{r}_{i,j}-\mathbf{r}_{i,j-1}\right|_{\mathrm{MIC}},
\end{equation}

where MIC denotes the use of minimum image convention and $n_a$ is the number of atoms. 

The average displacement $\bar{D_k}$ over all cells for an elapsed time of $\Delta t_k$ is: 

\begin{equation}
    \bar{D_k} = \frac{1}{20} \sum_j \bar{d_j}.
\end{equation}

Since the NpT ensemble was used, successive cells did not necessarily have the same volume. The cells were aligned at the origin and, in order to avoid having atoms outside of the cell, the cell parameters used for MIC displacement calculations were:

\begin{equation}
    a = \max(a_j, a_{j+1}), \quad b = \max(b_{j}, b_{j+1}), \quad c = \max(c_j, c_{j+1}).
\end{equation}

The average cell vectors mismatch was 0.37 , too little to significantly impact the onset of $\bar{D}$ saturation.

\begin{figure}[H]
    \centering
    \includegraphics[width=\linewidth]{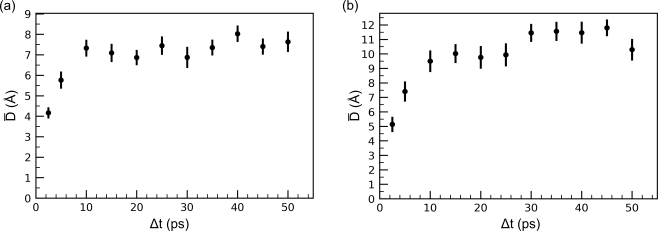}
    \caption{Average displacement of atoms as function of time elapsed between successive melt snapshots at 5000 K for (a) 192-atom cells, with an average volume of 3371 \AA{}$^3$,  (b) 648-atom cells, with an average volume of 11461 \AA{}$^3$ . The error bars give the standard error of the mean.}
    \label{fig:S1}
\end{figure}

A duration of 10 ps between snapshots was used for 192-atom simulations, while 40 ps was used for 648-atom simulations. Figure~\ref{fig:S1} shows that both values lead to saturated $\bar{D}$, thus maximising structural decorrelation.

\subsubsection{PBE BFGS cell optimisation convergence criteria}
\begin{itemize}
    \item Maximum step size = 0.003 (bohr)
    \item Root mean squared step size = 0.0015 (bohr)
    \item Maximum gradient 0.00045 (Eh/bohr)
    \item Root mean squared gradient = 0.0003 (Eh/bohr)
    \item Pressure deviation = 100 (bar)
\end{itemize}

\newpage
\subsubsection{Coordination cutoff choice}
The radial distribution function of the 2512 PBE optimised models has a broad minimum after the first sharp peak, extending between 1.7 \AA{} and 2.5 \AA{}. The sensitivity of coordination to the coordination cutoff was tested over the whole width of this broad minimum, with a step of 0.1 \AA{}. A cutoff of 1.9 \AA{} minimised coordination defects, but still miscategorised some Si$_5$ defects as Si$_4$ sites. Therefore, a conservative cutoff of 2.4 \AA{} was selected to ensure that all coordination defects were correctly identified and removed. 

\begin{figure}[H]
    \centering
    \includegraphics[width=0.5\linewidth]{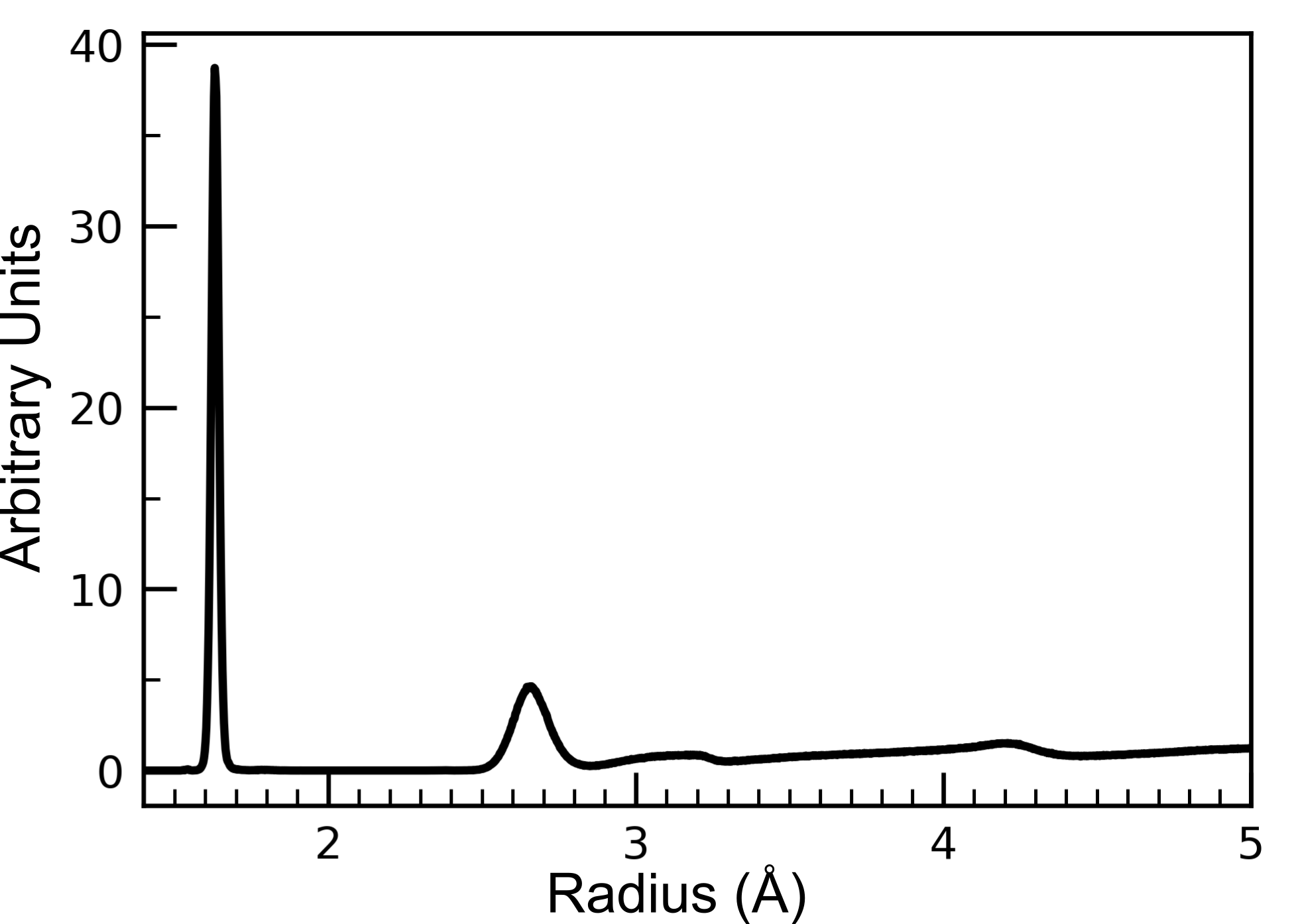}
    \caption{Total radial distribution function averaged over the 2512 PBE optimised models.}
    \label{fig:all_PBE_RDF}
\end{figure}

{\setlength{\tabcolsep}{7pt}
\begin{table}[H]
\centering
\caption{\label{tab:cutoff_coord_sensitivity}
Sensitivity of coordination to cutoff value}
\begin{tabular}{c|cccc|ccccc|c}
\toprule
Cutoff & O$_0$ & O$_1$ & O$_2$ & O$_3$ & Si$_1$ & Si$_2$ & Si$_3$ & Si$_4$ & Si$_5$ & \% Correct\\
\midrule
1.7 & 0.190 & 0.487 & 99.322 & 0.000 & 0.122 & 1.342 & 98.487 & 0.000 & 0.000 & 99.044 \\
1.8 & 0.014 & 0.247 & 99.713 & 0.026 & 0.001 & 0.515 & 99.463 & 0.021 & 0.000 & 99.630 \\
1.9 & 0.000 & 0.167 & 99.669 & 0.164 & 0.000 & 0.072 & 99.861 & 0.067 & 0.000 & 99.733 \\
2.0 & 0.000 & 0.165 & 99.646 & 0.188 & 0.000 & 0.033 & 99.887 & 0.080 & 0.000 & 99.727 \\
2.1 & 0.000 & 0.165 & 99.636 & 0.199 & 0.000 & 0.014 & 99.903 & 0.083 & 0.000 & 99.725 \\
2.2 & 0.000 & 0.165 & 99.631 & 0.204 & 0.000 & 0.008 & 99.905 & 0.087 & 0.000 & 99.723 \\
2.3 & 0.000 & 0.165 & 99.627 & 0.208 & 0.000 & 0.005 & 99.903 & 0.092 & 0.000 & 99.719 \\
2.4 & 0.000 & 0.164 & 99.626 & 0.210 & 0.000 & 0.004 & 99.901 & 0.095 & 0.000 & 99.718 \\
2.5 & 0.000 & 0.164 & 99.622 & 0.214 & 0.000 & 0.002 & 99.896 & 0.101 & 0.000 & 99.713 \\
\bottomrule
\end{tabular}
\end{table}}

\subsubsection{Structural properties of the defect-free PBE optimised models}

\begin{figure}[H]
    \centering
    \includegraphics[width=1\linewidth]{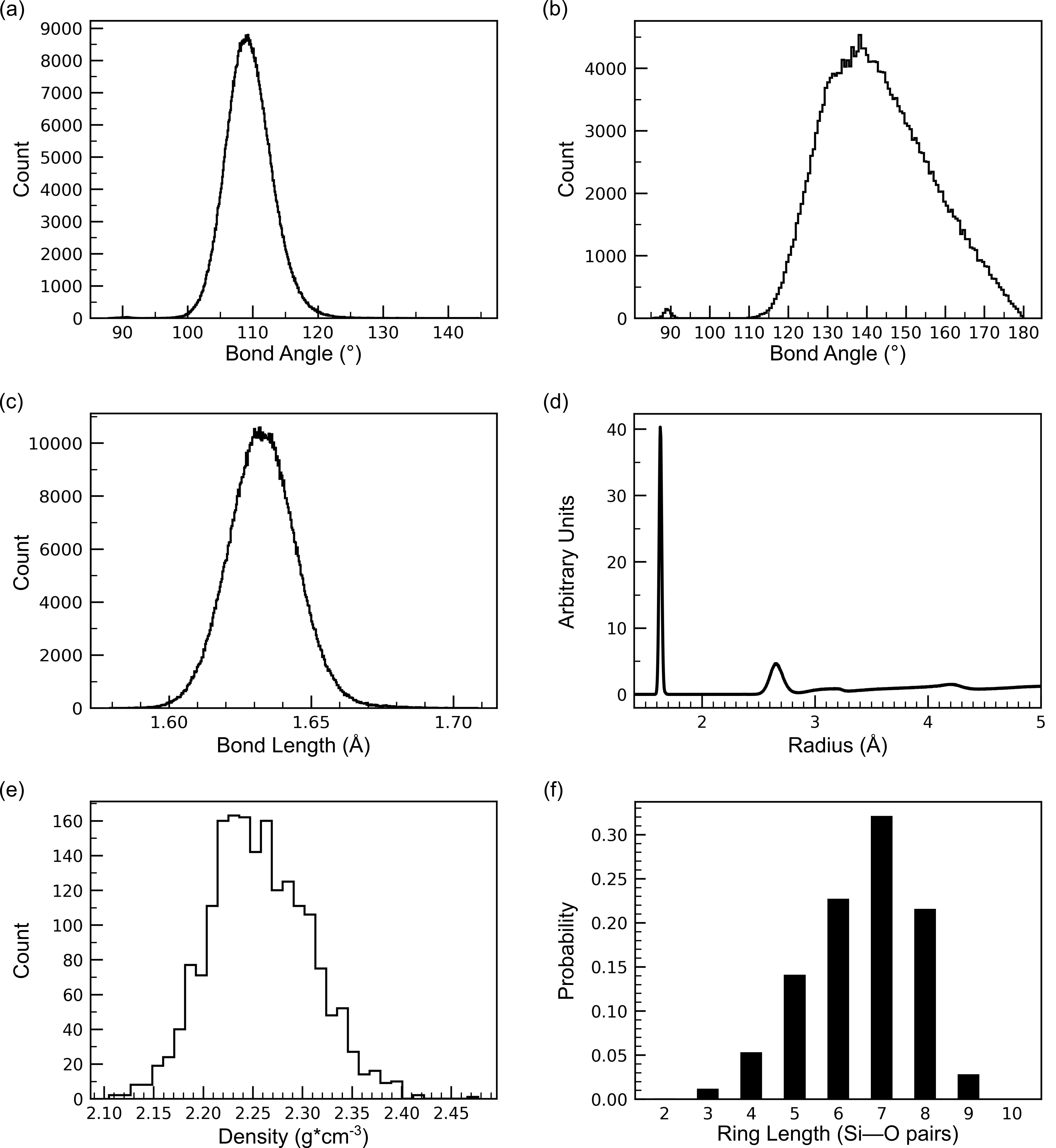}
    \caption{(a) \OSiO{} angle distribution, (b) \SiOSi{} angle distribution, (c) \SiO{} bond length distribution, (d) Total radial distribution function, (e) Density distribution, (f) Ring length distribution. The probabilities of 2- and 10-membered rings are 0.0008 and 0.0005 respectively, making them imperceptible on the chart.}
    \label{fig:structure_distributions}
\end{figure}

\newpage
\subsection{Structural descriptors of the neutral network}
\subsubsection{Ring Topology}
\begin{figure}[htb!]
    \centering
    \includegraphics[width=1\linewidth]{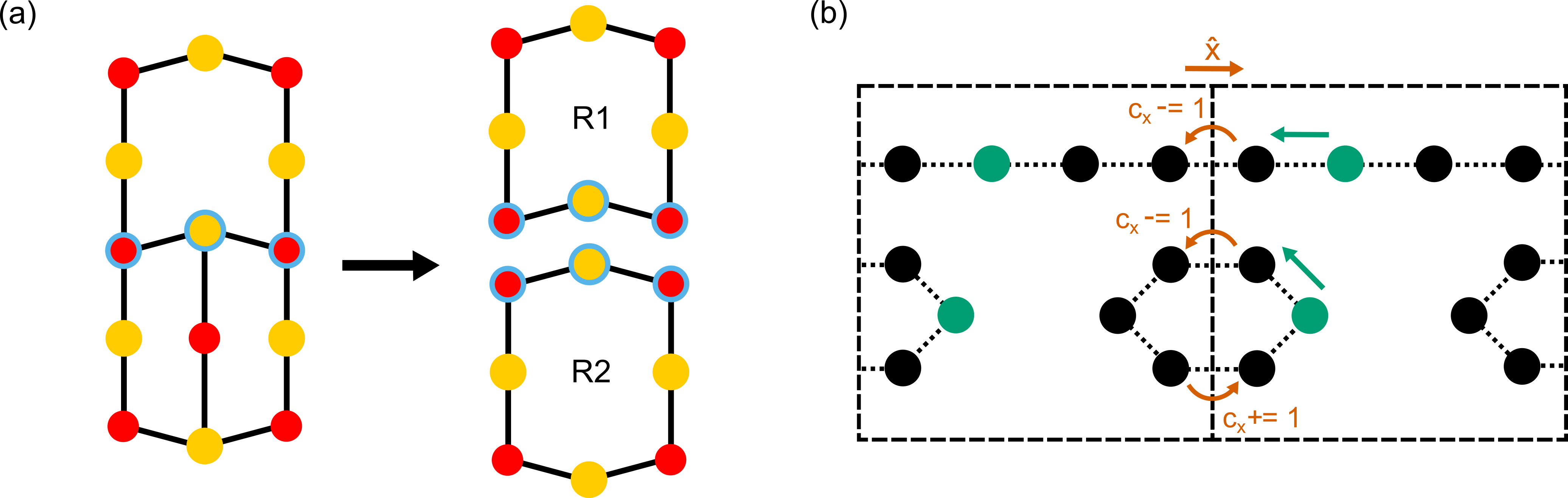}
    \caption{Ring selection rules: (a) Given an angle (highlighted with blue outlines), a ring is defined as the shortest closed loop containing the two bonds that define the angle. If more than one ring of the shortest length exist, they are all counted as valid rings. (b) In a periodic system, loops with a non-zero net translation in any cell direction (top) are periodic strings and are discarded. Loops with zero net translation in all cell directions (bottom) are valid rings.}
    \label{fig:rings_def}
\end{figure}

Rings were identified using the rules illustrated in Fig.~\ref{fig:rings_def}. All \OSiO{} and \SiOSi{} angles were assigned at least one ring. The procedure was as follows:\\
1) a neighbourlist was generated with a neighbour cut-off of 2.40 \r{A} for Si-O pairs and 0 \r{A} for O-O and Si-Si pairs.\\ 
2) The neighbourlist was converted into a graph using the python NetworkX library \cite{Hagberghagberg2008ExploringNetworkX}, each atom becoming a node and each bond an edge.\\
3) NetworkX's simple\_cycles algorithm was applied with a length cutoff of 20 nodes, generating a list of all closed loops of up to 20 nodes in the model.\\ 
4) Periodic strings were removed by keeping only the cycles with zero net translation in all cell directions, as illustrated in Fig.~\ref{fig:rings_def}(b).\\ 
5) The remaining cycles were filtered by keeping only the shortest rings for each O-Si-O and Si-O-Si angle in the model. If the angle contained more than one loop of the same minimum length, all such loops were counted. The resulting ring distribution of the 1865 defect-free bulk \aSiOtwo{} models is given in Fig.~\ref{fig:structure_distributions}(f).

\subsection{Ensuring statistical significance}

For the convergence of the \SiOfour{} packing distribution, the Wasserstein 2-distance (WS2D) was computed using the Python Optimal Transport library \cite{flamary2021pot,flamary_2025_17176920}. The distance matrix was obtained using the 'dist' function with the 'sqeuclidian' metric, and the earth mover's distance was calculated using the 'emd2' function with uniform weights.

\begin{figure}[htbp]
    \centering
    \includegraphics[width=1\linewidth]{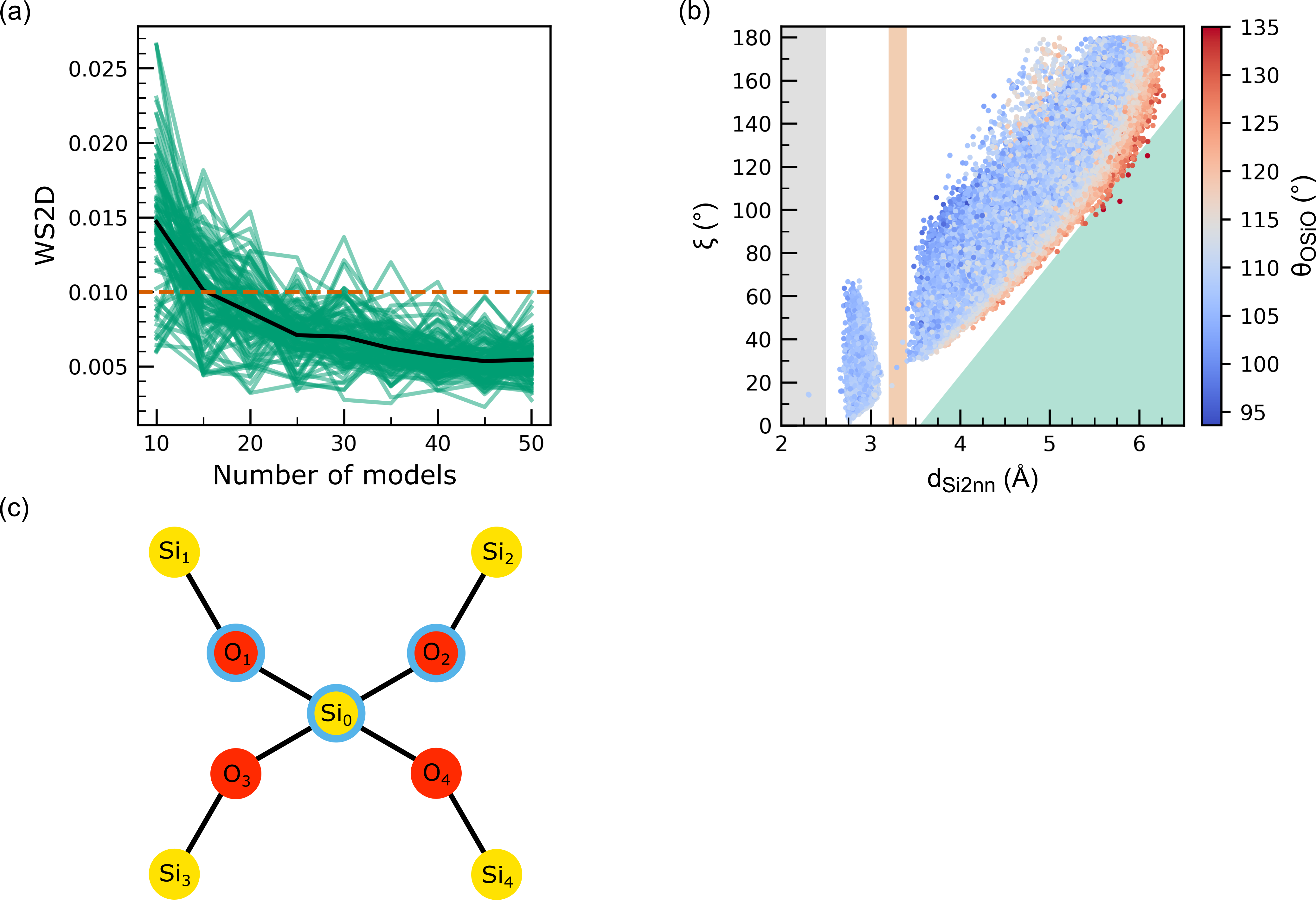}
    \caption{(a) Convergence of the global packing distribution in batches of 5 models. The orange dashed line indicates the convergence threshold WS2D=0.01. Each green line is the convergence curve for a different sample of 50 models picked randomly from the set of 1865 defect-free models. The black curve gives the average of the green curves. (b) Definition of the outlier regions of the packing distribution: $d_{\mathrm{Si2nn}} < 2.5$ (grey); $d_{\mathrm{Si2nn}} \in{} [3.2,3.4]$ (orange); $\xi < (51.5 \times d_{\mathrm{Si2nn}}-182.5)$ (green). (c) Schematic of a cluster of atoms surrounding an \OSiO{} site (highlighted with blue outlines). }
    \label{fig:S4_2nn_MRO}
\end{figure}

Figure~\ref{fig:S4_2nn_MRO}(a) shows the convergence of the global packing distribution of the 1865 defect-free PBE-optimised models. Due to the memory cost of the calculation, the convergence could not be calculated using all 1865 models. Instead, an average convergence was calculated using 100 samples of 50 randomly selected models. For each sample, the convergence was calculated in batches of 5 models by calculating the WS2D relative to the previous iteration. The two dimensions of the packing distribution were min-max normalised to give them equal weight in the convergence process and to make the WS2D unitless. The global packing distribution was found to be reliably converged within 50 models. 

Prior to downsampling, the following descriptors were defined to identify outlier models which were expected to lead to different electron-trapping behaviour:
\begin{itemize}
    \item $\theta_{\mathrm{OSiO,max}}^{\mathrm{model}}$: the largest \OSiO{} angle in a model. 
    \item \SiTwonnmaxmodel{}: the largest $d_{\mathrm{Si2nn}}$ in a model.
    \item $\lambda_{\mathrm{max}}^{\mathrm{model}}$: the largest value $\lambda$ in the sample, defined as the euclidean distance from the origin in the parameter space spanned by $d_{\mathrm{Si2nn}}$ and the \OSiO{} angle value.
\end{itemize}

During downsampling, the inclusion of the following models was manually enforced to test the electron trapping behaviour of :
\begin{itemize}
    \item the 5 models containing the largest $\theta_{\mathrm{OSiO,max}}^{\mathrm{model}}$.
    \item the 5 models containing the lowest $\theta_{\mathrm{OSiO,max}}^{\mathrm{model}}$.
    \item the 5 models containing the largest \SiTwonnmaxmodel{}.
    \item the 5 models containing the lowest \SiTwonnmaxmodel{}.
    \item the 5 models containing the largest $\lambda_{\mathrm{max}}^{\mathrm{model}}$.
    \item the 5 models containing the lowest $\lambda_{\mathrm{max}}^{\mathrm{model}}$.
    \item the models containing points in the outlier regions of the packing distribution as defined in Fig.~\ref{fig:S4_2nn_MRO}(b).
\end{itemize}

\newpage

In addition, the 20 samples used for test HSE06 electron trapping simulations were also manually included to conserve computational resources.

The total number of manually included models was 66, leaving 144 models to be randomly selected. The Kullback-Leibler divergence of the final 200 model subset's packing distribution relative to that of the full 1865 defect-free model set was 0.01.

\subsection{Hybrid-DFT electron-trapping calculation}
The following convergence criteria were used for both the neutral cell optimisations and the charged geometry optimisations:
\begin{itemize}
    \item Maximum step size = 0.003 (bohr)
    \item Root mean squared step size = 0.0015 (bohr)
    \item Maximum gradient 0.00045 (Eh/bohr)
    \item Root mean squared gradient = 0.0003 (Eh/bohr)
    \item Pressure deviation = 100 (bar)
\end{itemize}

\subsection{Precursor-site scoring}

A precursor site was defined as an O--Si--O angle in the neutral defect-free model on which an injected electron localises after relaxation of the $q=-1$ cell.  

For each candidate site $i$, a set of $N$ structural descriptors $x_{ij}$ was calculated from the neutral geometry. The score of site i is defined as:
\begin{equation}
    s_i=\sum_j w_j z_{ij}, 
    \qquad
    z_{ij}=(x_{ij}-\mu_j)/\sigma_j,
    \label{eq:precursor_score_current}
\end{equation}
where $\mu_j$ and $\sigma_j$ are respectively the mean and standard deviation of descriptor $j$ over the set of 200 neutral HSE06 models, and $\omega_j$ is a corresponding weight between 0 and 1. The score is constructed so that larger values indicate sites more favourable for electron trapping.

The precursor score was then used to rank sites within each model. The site with the largest score was assigned a rank of 1 and the site with lowest score had a rank of 384. A rank-1 prediction occurs when the actual precursor site has the largest score in its cell.

Thirty-six candidate descriptors of short and medium range order were considered for the precursor score. Given an angle $\mathrm{O_1}$--$\mathrm{Si_0}$--$\mathrm{O_2}$, highlighted in blue in Fig.~\ref{fig:S4_2nn_MRO}(c), the list of candidate descriptors for the precursor score considered in this work was as follows:
\newline

\begin{description}
    \item[$R_l^{\mathrm{min}}$]
    Length of the smallest ring that $\angle\mathrm{O_1Si_0O_2}$ is a part of,
    \item[$V_{\mathrm{Si_0}}$]
    Voronoi volume associated with atom Si$_0$
    \item[$V_{\mathrm{O}}^{\mathrm{min}}$] 
    \begin{equation}
        V_{\mathrm{O}}^{\mathrm{min}} = \min\left(V_{\mathrm{O_1}}, V_{\mathrm{O_2}}\right),
    \end{equation}
    \item[$V_{\mathrm{O}}^{\mathrm{avrg}}$]
    \begin{equation}
        V_{\mathrm{O}}^{\mathrm{avrg}} = \frac{V_{\mathrm{O_1}}+V_{\mathrm{O_2}}}{2},
    \end{equation}
    \item[$V_{\mathrm{O}}^{\mathrm{max}}$]
    \begin{equation}
        V_{\mathrm{O}}^{\mathrm{max}} = \max\left(V_{\mathrm{O_1}}, V_{\mathrm{O_2}}\right),
    \end{equation}
    \item[$V_T$]
    \begin{equation}
        V_T = V_{\mathrm{Si_0}}+V_{\mathrm{O_1}}+V_{\mathrm{O_2}},
    \end{equation}
    \item[$d_{\mathrm{Sinn}}^{\mathrm{min}}$]
    \begin{equation}
        d_{\mathrm{Sinn}}^{\mathrm{min}} = \min(\left|\mathbf{r}_{\mathrm{Si}_1}-\mathbf{r}_{\mathrm{Si}_0}\right|_{\mathrm{MIC}},\left|\mathbf{r}_{\mathrm{Si}_2}-\mathbf{r}_{\mathrm{Si}_0}\right|_{\mathrm{MIC}}),
    \end{equation}
    \item[$d_{\mathrm{Sinn}}^{\mathrm{avrg}}$]
    \begin{equation}
        d_{\mathrm{Sinn}}^{\mathrm{avrg}} = \frac{(\left|\mathbf{r}_{\mathrm{Si}_1}-\mathbf{r}_{\mathrm{Si}_0}\right|_{\mathrm{MIC}}+\left|\mathbf{r}_{\mathrm{Si}_2}-\mathbf{r}_{\mathrm{Si}_0}\right|_{\mathrm{MIC}})}{2},
    \end{equation}
    \item[$d_{\mathrm{Sinn}}^{\mathrm{max}}$]
    \begin{equation}
        d_{\mathrm{Sinn}}^{\mathrm{max}} = \max(\left|\mathbf{r}_{\mathrm{Si}_1}-\mathbf{r}_{\mathrm{Si}_0}\right|_{\mathrm{MIC}},\left|\mathbf{r}_{\mathrm{Si}_2}-\mathbf{r}_{\mathrm{Si}_0}\right|_{\mathrm{MIC}}),
    \end{equation}
    \item[$d_{\mathrm{Si-O}}^{\mathrm{avrg}}$]
    \begin{equation}
        d_{\mathrm{Si-O}}^{\mathrm{avrg}} = \frac{(\left|\mathbf{r}_{\mathrm{O}_1}-\mathbf{r}_{\mathrm{Si}_0}\right|_{\mathrm{MIC}}+\left|\mathbf{r}_{\mathrm{O}_2}-\mathbf{r}_{\mathrm{Si}_0}\right|_{\mathrm{MIC}})}{2},
    \end{equation}
    \item[$d_{\mathrm{Si-O}}^{\mathrm{max}}$]
    \begin{equation}
        d_{\mathrm{Si-O}}^{\mathrm{max}} = \max(\left|\mathbf{r}_{\mathrm{O}_1}-\mathbf{r}_{\mathrm{Si}_0}\right|_{\mathrm{MIC}},\left|\mathbf{r}_{\mathrm{O}_2}-\mathbf{r}_{\mathrm{Si}_0}\right|_{\mathrm{MIC}}),
    \end{equation}
    \item[$\theta_{\mathrm{SiOSi}}^{\mathrm{min}}$]
    \begin{equation}
        \theta_{\mathrm{SiOSi}}^{\mathrm{min}} = \min(\angle \mathrm{Si_1O_1Si_0},\angle\mathrm{Si_2O_2Si_0}),
    \end{equation}
    \item[$\theta_{\mathrm{SiOSi}}^{\mathrm{avrg}}$]
    \begin{equation}
        \theta_{\mathrm{SiOSi}}^{\mathrm{avrg}} = \frac{\angle \mathrm{Si_1O_1Si_0}+\angle\mathrm{Si_2O_2Si_0}}{2},
    \end{equation}
    \item[$\theta_{\mathrm{SiOSi}}^{\mathrm{max}}$]
    \begin{equation}
        \theta_{\mathrm{SiOSi}}^{\mathrm{max}} = \max(\angle \mathrm{Si_1O_1Si_0},\angle\mathrm{Si_2O_2Si_0}),
    \end{equation}
    \item[$d_{\mathrm{Si2nn}}$]
    \begin{equation}
        d_{\mathrm{Si2nn}} = \left|\mathbf{r}_{\mathrm{Si}_1}-\mathbf{r}_{\mathrm{Si}_2}\right|_{\mathrm{MIC}},
    \end{equation}
    \item[$\theta_{\mathrm{OSiO}}$]
    \begin{equation}
        \theta_{\mathrm{OSiO}} = \angle\mathrm{O_1Si_0O_2}
    \end{equation}
    \item[$\theta_{\mathrm{SiSiSi}}$]
    \begin{equation}
        \theta_{\mathrm{SiSiSi}} = \angle\mathrm{Si_1Si_0Si_2}
    \end{equation}
    \item[$\xi$]
    \begin{equation}
        \xi = \cos^{-1}\left(\hat{\mathbf{u}}_1 \cdot \hat{\mathbf{u}}_2\right),
    \end{equation}
    with
    \begin{equation}
        \hat{\mathbf{u}}_1 =
    \frac{\mathbf{r}_{\mathrm{O}_1}-\mathbf{r}_{\mathrm{Si}_1}}
    {\left|\mathbf{r}_{\mathrm{O}_1}-\mathbf{r}_{\mathrm{Si}_1}\right|},
    \qquad
    \hat{\mathbf{u}}_2 =
    \frac{\mathbf{r}_{\mathrm{O}_2}-\mathbf{r}_{\mathrm{Si}_2}}
    {\left|\mathbf{r}_{\mathrm{O}_2}-\mathbf{r}_{\mathrm{Si}_2}\right|}.
    \end{equation}
\end{description}

Additionally, the equivalents of all the descriptors listed above for the opposing angle, i.e. $\angle\mathrm{O_3Si_0O_4}$, were also candidates for the precursor score of $\angle\mathrm{O_1Si_0O_2}$. These are designated by the 'opp-' prefix in the tables bellow, e.g.:
\begin{equation}
    \mathrm{opp-}\theta_{\mathrm{OSiO}} = \angle\mathrm{O_3Si_0O_4}
\end{equation}

The procedure to build the best performing precursor score was as follows:\\

1) The sensitivity of each descriptor to trap precursor sites was measured using the average trap site rank if that descriptor alone was used as the score.\\

2) A correlation matrix of all descriptors was built using the Scipy pearsonr metric \cite{Virtanen2020SciPyPython}. All pairs of descriptors with a Pearson coefficient larger than 0.8 were identified and, starting with the pair with the highest correlation, the descriptor with the lowest sensitivity was removed.\\

3) The 10 descriptors with the highest sensitivities (listed in table \ref{tab:descriptor_sensitivity}) were selected, and the precursor score was calculated for every possible combination of descriptors, with equal weights of 1.\\ 

4) The 10 combinations that correctly predicted the most precursor sites (by assigning them a rank of 1) were selected for weight optimisation. This was done using a grid search, with the weights taking values from 0 to 1 in steps of 0.1. The optimisation results are listed in table \ref{tab:weight_opt} \\

5) The set of descriptors with optimised weights that correctly predicted the most precursor sites was selected as the final precursor score definition. 

\begin{table}[H]
\centering
\caption{\label{tab:descriptor_sensitivity}
Top 10 most IET-sensitive descriptors after removal of redundant descriptors}
\begin{tabular}{lc}
\toprule
Descriptor & Average Trap Site Rank \\
\midrule
$\theta_{\mathrm{OSiO}}$ & 10 \\
\midrule
$d_{\mathrm{Si-O}}^{\mathrm{avrg}}$ & 21 \\
\midrule
$V_{\mathrm{Si_0}}$ & 32 \\
\midrule
$d_{\mathrm{Si2nn}}$ & 43 \\
\midrule
$R_l^{\mathrm{min}}$ & 59 \\
\midrule
$\mathrm{opp-}\theta_{\mathrm{OSiO}}$ & 129 \\
\midrule
$d_{\mathrm{Sinn}}^{\mathrm{max}}$ & 135 \\
\midrule
$\mathrm{opp-}R_l^{\mathrm{min}}$ & 138 \\
\midrule
$\mathrm{opp-}d_{\mathrm{Si-O}}^{\mathrm{max}}$ & 139 \\
\midrule
$\mathrm{opp-}\theta_{\mathrm{SiSiSi}}$ & 170 \\
\bottomrule
\end{tabular}
\end{table}

\clearpage
\begin{turnpage}
\begin{table}[H]
\caption{\label{tab:weight_opt} Weight optimisation of the top-10 best performing descriptor combinations}
\begin{ruledtabular}
\begin{tabular}{p{0.42\textwidth} c p{0.3\textwidth} c}
Descriptor combination &
\parbox[t]{2.6cm}{\centering unoptimised rank-1 prediction count} &
Optimised weights &
\parbox[t]{2.6cm}{\centering optimised rank-1 prediction count} \\
\midrule
$\theta_{\mathrm{OSiO}}$, $d_{\mathrm{Si-O}}^{\mathrm{avrg}}$, $V_{\mathrm{Si_0}}$, $d_{\mathrm{Si2nn}}$, $d_{\mathrm{Sinn}}^{\mathrm{max}}$ & 99 & 0.7; 0.3; 0.2; 0.5; 0.4 & 104 \\
\midrule
$\theta_{\mathrm{OSiO}}$, $d_{\mathrm{Si-O}}^{\mathrm{avrg}}$, $V_{\mathrm{Si_0}}$, $d_{\mathrm{Sinn}}^{\mathrm{max}}$ & 94 & 0.9; 0.5; 0.3; 0.7 & 97 \\
\midrule
$\theta_{\mathrm{OSiO}}$, $d_{\mathrm{Si-O}}^{\mathrm{avrg}}$, $d_{\mathrm{Si2nn}}$, $d_{\mathrm{Sinn}}^{\mathrm{max}}$ & 93 & 0.8; 0.6; 0.6; 0.3 & 99 \\
\midrule
$\theta_{\mathrm{OSiO}}$, $d_{\mathrm{Si-O}}^{\mathrm{avrg}}$, $V_{\mathrm{Si_0}}$ & 89 & 0.9; 1.0; 0.9 & 91 \\
\midrule
$\theta_{\mathrm{OSiO}}$, $d_{\mathrm{Si-O}}^{\mathrm{avrg}}$, $V_{\mathrm{Si_0}}$, $d_{\mathrm{Si2nn}}$, $R_l^{\mathrm{min}}$, $d_{\mathrm{Sinn}}^{\mathrm{max}}$ & 89 & 0.7; 0.3; 0.2; 0.5; 0.0; 0.4 & 104 \\
\midrule
$\theta_{\mathrm{OSiO}}$, $d_{\mathrm{Si-O}}^{\mathrm{avrg}}$, $V_{\mathrm{Si_0}}$, $R_l^{\mathrm{min}}$, $d_{\mathrm{Sinn}}^{\mathrm{max}}$ & 88 & 0.9; 0.5; 0.3; 0.5; 0.7 & 101 \\
\midrule
$\theta_{\mathrm{OSiO}}$, $d_{\mathrm{Si-O}}^{\mathrm{avrg}}$, $V_{\mathrm{Si_0}}$, $R_l^{\mathrm{min}}$ & 87 & 0.9; 1.0; 0.9; 0.1 & 92 \\
\midrule
$\theta_{\mathrm{OSiO}}$, $d_{\mathrm{Si-O}}^{\mathrm{avrg}}$, $d_{\mathrm{Sinn}}^{\mathrm{max}}$ & 87 & 0.8; 0.6; 0.7 & 91 \\
\midrule
$\theta_{\mathrm{OSiO}}$, $d_{\mathrm{Si-O}}^{\mathrm{avrg}}$, $V_{\mathrm{Si_0}}$, $d_{\mathrm{Si2nn}}$ & 87 & 0.6; 0.8; 0.1; 1.0 & 94 \\
\midrule
$\theta_{\mathrm{OSiO}}$, $d_{\mathrm{Si-O}}^{\mathrm{avrg}}$, $d_{\mathrm{Si2nn}}$ & 86 & 0.3; 0.4; 0.3 & 90 \\
\bottomrule
\end{tabular}
\end{ruledtabular}
\end{table}
\end{turnpage}

\clearpage
\section{Results Supplementary Information}
\subsection{One- to two-electron trap evolution}
\begin{table}[H]
\centering
\caption{\label{tab:polaron_to_bipolaron_evol}
Evolution from the trapped polaron relaxed state to the final trapped bipolaron relaxed state. ``NT'' denotes cells in which the electron or electrons occupied conduction-band-like states.}
\begin{tabular}{lll}
\toprule
Initial $1e^-$ outcome & Final $2e^-$ motif & No. of cells \\
\midrule
Ed & bi-Si$_3$ & 121 \\
F & bi-Si$_3$ & 7 \\
NT & bi-Si$_3$ & 7 \\
Si$_3$ & bi-Si$_3$ & 1 \\
\midrule
Ed & bi-Si$_3$--Ed & 6 \\
Ed & bi-Si$_3$--F & 6 \\
F & bi-Si$_3$--F & 1 \\
\midrule
Ed & bi-Ed & 6 \\
\midrule
Ed & bi-Ed--Ed & 15 \\
NT & bi-Ed--Ed & 1 \\
Ed & bi-Ed--F & 20 \\
NT & bi-Ed--F & 1 \\
F & bi-F--F & 1 \\
F--F & bi-F--F & 1 \\
NT & bi-F--F & 1 \\
\midrule
NT & NT & 5 \\
\bottomrule
\end{tabular}
\end{table}

\subsection{Distributions of descriptors included in the precursor score}

\begin{figure}[H]
    \centering
    \includegraphics[width=0.9\linewidth]{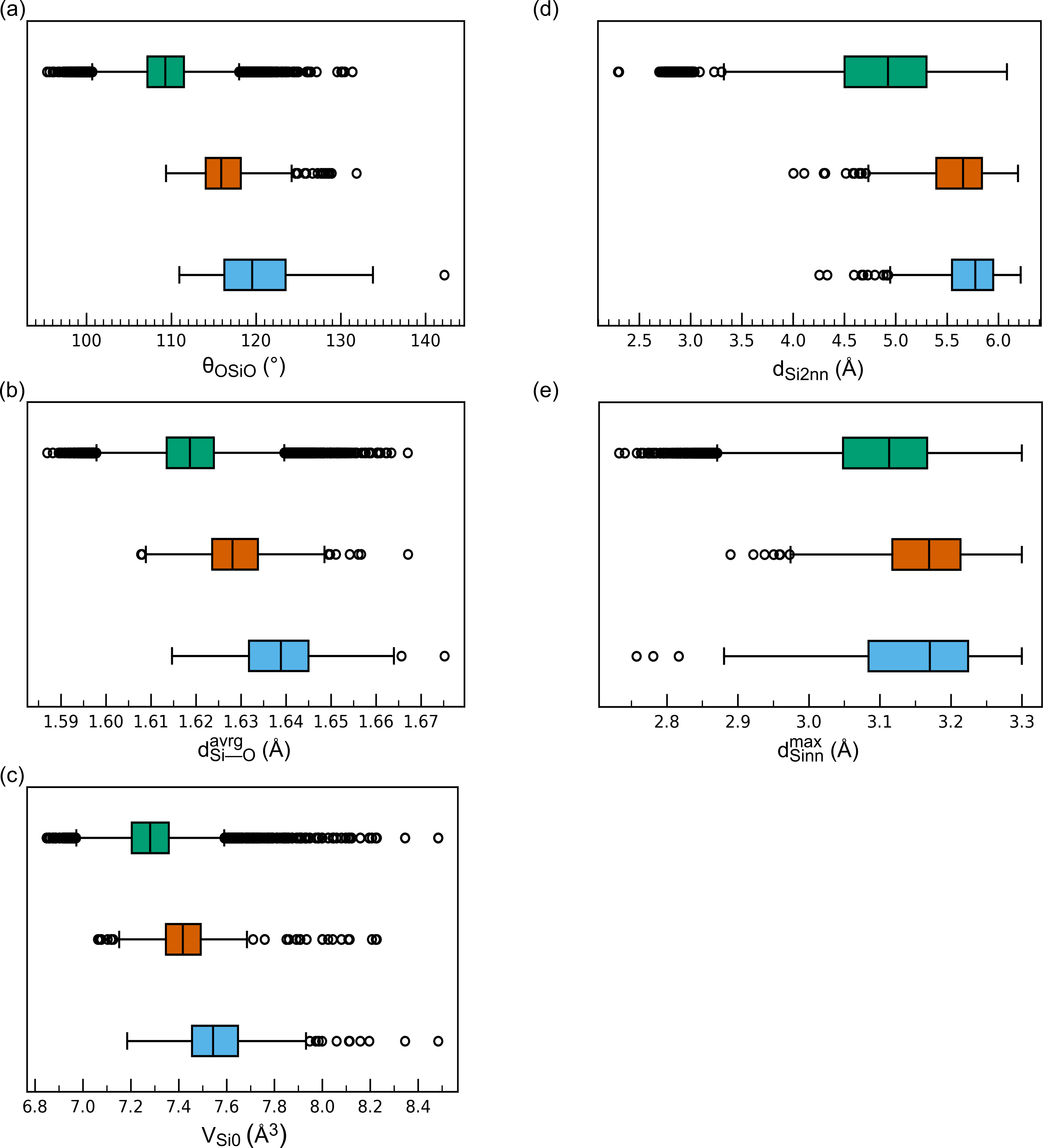}
    \caption{Distributions of descriptors used in the composite precursor score. Blue boxes correspond to Ed precursor sites, orange boxes correspond to higher-ranked non-selected sites, and green boxes correspond to all other sites. Boxes span the first and third quartiles; a notch marks the median; whiskers extend to the nearest point within 1.5 times the interquartile range; black circles beyond the whiskers are outliers. (a) \OSiO{} angle, (b) average \SiO{} bond length, (c) Voronoi volume of the central Si atom, (d) Si second nearest-neighbour distance, (e) Maximum Si nearest-neighbour distance.}
    \label{fig:S5_precursor_characterisation}
\end{figure}

\newpage
\section{Discussion Supplementary Information}
\subsection{Tail density}
\begin{figure}[H]
    \centering
    \includegraphics[width=1\linewidth]{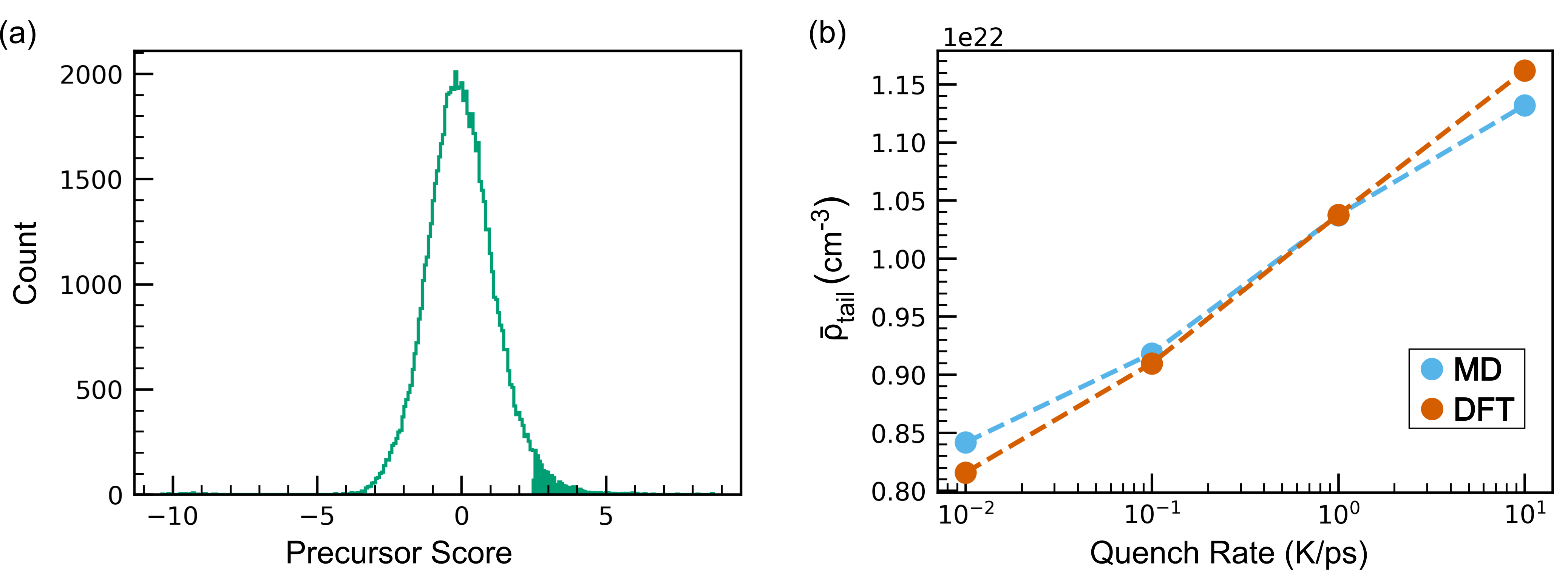}
    \caption{(a) Precursor score distribution of the HSE06 subset. The tail is shaded for values larger than 2.49. (b) Comparison of precursor score tail density for 192-atom cells immediately after MD and after DFT optimisation using PBE.}
    \label{fig:prec_score_tail_dens}
\end{figure}

Figure~\ref{fig:prec_score_tail_dens}(a) shows the distribution of precursor scores for all sites in the neutral HSE06 optimised models. The lowest score for an actual precursor site in the set was 2.49. This score served as a threshold to define the precursor tail density $\bar\rho_{\mathrm{tail}}$. Figure~\ref{fig:prec_score_tail_dens}(b) compares the precursor tail density immediately after MD and after PBE optimisation for 192-atom models at different quench rates. Each data point was calculated based on at least 89 models. There is little difference is $\bar\rho_{\mathrm{tail}}$ between the two levels of optimisation. Therefore, only MD results were used when comparing with 648-atom cells in the main text.

\newpage
\subsection{Relationship between ring size, packing, and electron trapping}

\begin{figure}[H]
    \centering
    \includegraphics[width=\linewidth]{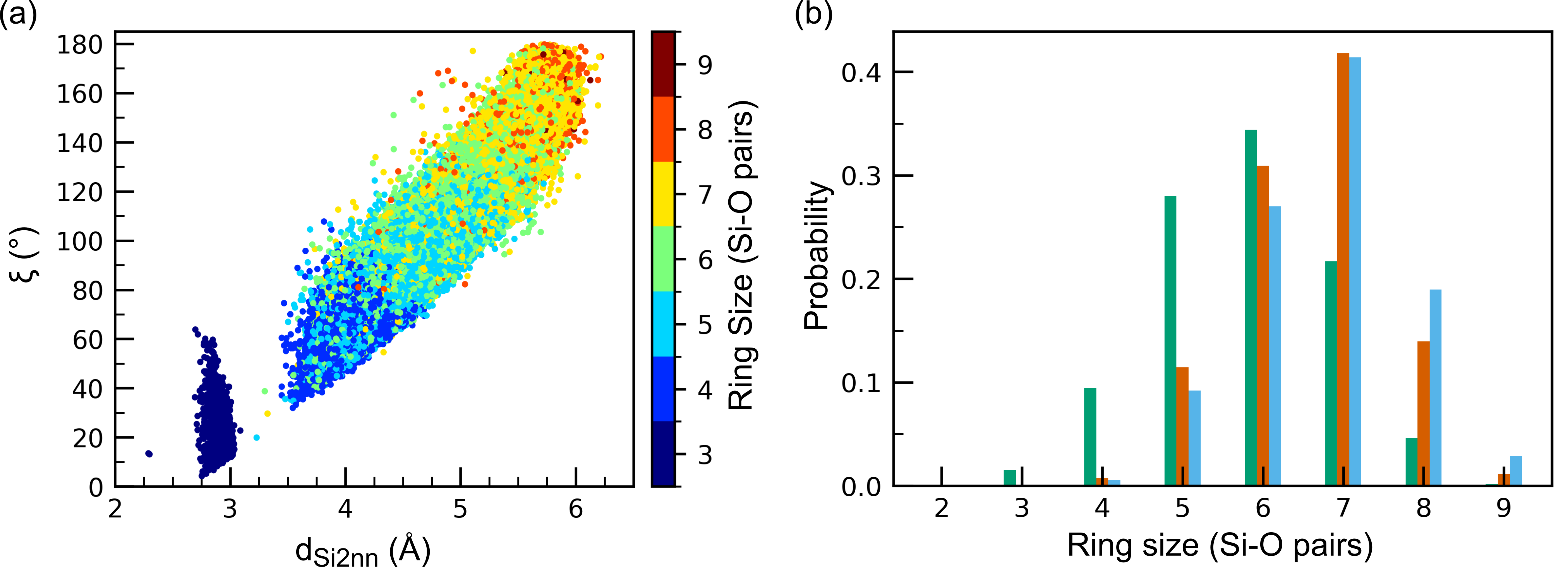}
    \caption{Relationship between ring length, packing geometry and trapping propensity. (a) Scatter-plot of the packing distribution of the 200-model. Each point is an \OSiO{} site. The colour indicates the size of the smallest ring that the site is a part of. (b) Distribution of the minimum ring lengths for Ed IET precursor sites (blue), non-selected higher-ranked sites (orange) and all other sites (green).}
    \label{fig:S8_Packing_vs_Rings}
\end{figure}

Smaller rings in \aSiOtwo{} have narrow inflexible \OSiO{} angles, associated with more tightly spread $d_{\mathrm{Si2nn}}$ values, as seen for three-membered rings in \figref{fig:S8_Packing_vs_Rings}(a). Values of $\xi$ remain broadly distributed for all ring sizes, and are associated with the degree of planarity of neighbouring sites. For example, planar three-membered rings have $\xi$ values that approach 0$^\circ$, while twisted three-membered rings can have $\xi$ values larger than 60$^\circ$. As the ring size increases, \OSiO{} angles can widen and the spread of $d_{\mathrm{Si2nn}}$ values increases, leading to a more flexible medium range environment. Consequently, larger rings are more favourable for electron trapping, as shown in \figref{fig:S8_Packing_vs_Rings}(b).

\newpage
\subsection{Spin treatment of two-electron cells}
All the energies listed in table \ref{tab:bipolaron_summary} are in eV. The S and T superscripts denote singlet and triplet energies respectively. $E^\mathrm{T}_{\mathrm{i1}}$ is defined as the average singlet bipolaron finite-size electrostatic error, while $E^\mathrm{T}_{\mathrm{i2}}$ is defined as two times the average polaron finite-size electrostatic error. The 'pa' subscript denotes potential alignment corrections while the 'uncor' subscript denotes uncorrected energies. The triplet potential alignment correction $E^\mathrm{T}_{\mathrm{pa}}$ is defined as the average singlet bipolaron potential alignment correction. The energy differences are:

\begin{equation}
    \Delta E_{\mathrm{uncor}} = E^\mathrm{T}_{\mathrm{uncor}} - E^\mathrm{S}_{\mathrm{uncor}},
\end{equation}
\begin{equation}
    \Delta E_1 = (E^\mathrm{T}_{\mathrm{uncor}}+E^\mathrm{T}_{\mathrm{i1}}+E^\mathrm{T}_{\mathrm{pa}})-(E^\mathrm{S}+E^\mathrm{S}_{\mathrm{i}}+E^\mathrm{S}_{\mathrm{pa}}),
\end{equation}
\begin{equation}
    \Delta E_2 = (E^\mathrm{T}_{\mathrm{uncor}}+E^\mathrm{T}_{\mathrm{i2}}+E^\mathrm{T}_{\mathrm{pa}})-(E^\mathrm{S}+E^\mathrm{S}_{\mathrm{i}}+E^\mathrm{S}_{\mathrm{pa}})
\end{equation}.\newline

\newpage
\footnotesize
\setlength{\LTpre}{0pt}
\setlength{\LTpost}{0pt}
\begin{longtable}{l|ccc|cccc|ccc}
\caption{\label{tab:bipolaron_summary} Charge correction of singlet and triplet bipolarons. All values are in eV.} \\
\toprule
Sample & $E^\mathrm{S}_{\mathrm{uncor}}$ & $E^\mathrm{S}_{\mathrm{i}}$ & $E^\mathrm{S}_{\mathrm{pa}}$ & $E^\mathrm{T}_{\mathrm{uncor}}$ & $E^{\mathrm{T}}_{\mathrm{i1}}$ & $E^{\mathrm{T}}_{\mathrm{i2}}$ & $E^\mathrm{T}_{\mathrm{pa}}$ & $\Delta E_{\mathrm{uncor}}$ & $\Delta E_{1}$ & $\Delta E_{2}$ \\
\midrule
1 & -62914.38 & 1.10 & 0.62 & -62913.36 & 1.09 & 0.55 & 0.43 & 1.02 & 0.81 & 0.26 \\
2 & -62918.28 & 1.08 & 0.67 & -62917.35 & 1.09 & 0.55 & 0.43 & 0.93 & 0.70 & 0.15 \\
3 & -62919.72 & 1.08 & 0.24 & -62918.11 & 1.09 & 0.55 & 0.43 & 1.60 & 1.81 & 1.26 \\
4 & -62918.26 & 1.08 & 0.24 & -62917.59 & 1.09 & 0.55 & 0.43 & 0.67 & 0.86 & 0.32 \\
5 & -62913.17 & 1.10 & 0.50 & -62912.39 & 1.09 & 0.55 & 0.43 & 0.78 & 0.70 & 0.15 \\
6 & -62916.58 & 1.09 & 0.28 & -62916.36 & 1.09 & 0.55 & 0.43 & 0.22 & 0.38 & -0.17 \\
7 & -62915.34 & 1.10 & -0.02 & -62912.58 & 1.09 & 0.55 & 0.43 & 2.76 & 3.20 & 2.65 \\
8 & -62916.64 & 1.09 & 0.21 & -62915.81 & 1.09 & 0.55 & 0.43 & 0.83 & 1.05 & 0.51 \\
9 & -62919.70 & 1.07 & 0.66 & -62919.02 & 1.09 & 0.55 & 0.43 & 0.69 & 0.48 & -0.07 \\
10 & -62915.83 & 1.09 & 0.35 & -62915.43 & 1.09 & 0.55 & 0.43 & 0.40 & 0.48 & -0.06 \\
11 & -62915.92 & 1.09 & 0.45 & -62915.35 & 1.09 & 0.55 & 0.43 & 0.57 & 0.55 & 0.01 \\
12 & -62918.17 & 1.08 & 0.83 & -62917.37 & 1.09 & 0.55 & 0.43 & 0.80 & 0.42 & -0.13 \\
13 & -62915.66 & 1.09 & 0.46 & -62915.52 & 1.09 & 0.55 & 0.43 & 0.14 & 0.11 & -0.43 \\
14 & -62916.12 & 1.09 & 0.30 & -62915.15 & 1.09 & 0.55 & 0.43 & 0.97 & 1.09 & 0.54 \\
15 & -62919.17 & 1.09 & 0.37 & -62916.79 & 1.09 & 0.55 & 0.43 & 2.38 & 2.45 & 1.90 \\
16 & -62919.28 & 1.09 & 0.66 & -62918.48 & 1.09 & 0.55 & 0.43 & 0.80 & 0.58 & 0.03 \\
17 & -62915.15 & 1.10 & 0.57 & -62914.36 & 1.09 & 0.55 & 0.43 & 0.78 & 0.64 & 0.09 \\
18 & -62919.14 & 1.08 & 0.44 & -62917.87 & 1.09 & 0.55 & 0.43 & 1.28 & 1.28 & 0.73 \\
19 & -62913.64 & 1.11 & 0.26 & -62912.42 & 1.09 & 0.55 & 0.43 & 1.23 & 1.38 & 0.83 \\
20 & -62919.90 & 1.08 & 0.77 & -62918.61 & 1.09 & 0.55 & 0.43 & 1.29 & 0.97 & 0.42 \\
\midrule
Averages & & & & & & & & 1.01 & 1.00 & 0.45 \\
\bottomrule
\end{longtable}

\newpage
\bibliographystyle{apsrev4-2}
\bibliography{references}